\documentclass[prb,twocolumn,showpacs,amsmath,amssymb,letterpaper,eqsecnum,floatfix]{revtex4}

\usepackage{color,graphicx}% Include figure files
\usepackage{wasysym}
\usepackage{times}

\newcommand{\rem}[1]{}
\newcommand{\SAVE}[1]{}
\newcommand{\LATER}[1]{}

\newcommand{\eqr}[1]{(\ref{#1})}
\newcommand{\epsSC}{{\varepsilon^*}_{\rm SC}}
\newcommand{\Hvec}{\mathbf{H}}
\newcommand{\Svec}{\mathbf{S}}
\newcommand{\Vvec}{\mathbf{V}}
\newcommand{\Wvec}{\mathbf{W}}
\newcommand{\Uvec}{\mathbf{U}}

\newcommand{\chivec}{\boldsymbol{\chi}}

\newcommand{\Mvec}{\mathbf{M}}
\newcommand{\uvec}{\mathbf{u}}

\newcommand{\vvec}{\mathbf{v}}

\newcommand{\Ovec}{\mathbf{0}}
\newcommand{\muvec}{\boldsymbol{\mu}}
\newcommand{\nuvec}{\boldsymbol{\nu}}
\newcommand{\etvec}{\boldsymbol{\eta}}
\newcommand{\svec}{\boldsymbol{\sigma}}
\newcommand{\xivec}{\vec{\xi}}
\newcommand{\qvec}{\mathbf{q}}
\newcommand{\Qvec}{\mathbf{Q}}

\newcommand{\rvec}{\mathbf{r}}
\newcommand{\Rvec}{\mathbf{R}}
\newcommand{\Dvec}{\boldsymbol{\Delta}}
\newcommand{\tJ}{{\tilde{J}}}
\newcommand{\ham}{\mathcal{H}}
\newcommand{\OO}{\mathcal{O}}
\newcommand{\LL}{\mathcal{L}}
\newcommand{\WW}{\mathcal{W}}
\newcommand{\PP}{\mathcal{P}}

\newcommand{\adag}{a^\dagger}

\newcommand{\tr}{\ensuremath{\operatorname{Tr}}}
\newcommand{\MF}{\mathrm{MF}}
\newcommand{\eff}{\mathrm{eff}}

\newcommand{\cl}{\mathrm{cl}}
\newcommand{\harm}{\mathrm{harm}}
\newcommand{\out}{\mathrm{out}}
\newcommand{\quart}{\mathrm{quart}}
\newcommand{\var}{\mathrm{var}}

\newcommand{\ord}{\mathrm{ord}}
\newcommand{\half}{\frac{1}{2}}

\begin{document}

\title{Anharmonic Ground state selection in the pyrochlore antiferromagnet}
\author{U. Hizi}
\altaffiliation{Current address: Samsung Semiconductor Israel Research and Design Center,
10 Oholiav St. Ramat Gan, 52522, Israel}
\author{C.~L.~Henley}
\email{clh@ccmr.cornell.edu}
\affiliation{Laboratory of Atomic and Solid State Physics, Cornell University, Ithaca, New York 14853-2501, USA}
\date{\today}

\begin{abstract} 
In the pyrochlore lattice Heisenberg antiferromagnet, for large spin length $S$,
the massive classical ground state degeneracy is partly lifted by the zero-point
energy of quantum fluctuations at harmonic order in spin-waves.
However, there remains an infinite manifold of degenerate collinear
ground states, related by a gaugelike symmetry.
We have extended the spin-wave calculation to quartic order, 
assuming a Gaussian variational wavefunction 
(equivalent to Hartree-Fock approximation). 
Quartic calculations \emph{do} break the harmonic-order degeneracy of periodic 
ground states. The form of the effective Hamiltonian describing this 
splitting, which depends on loops, was fitted numerically and also
rationalized analytically.  We find a family of states that are still 
almost degenerate, being split by the term from loops of length 26.
We also calculated the anharmonic terms for the checkerboard lattice,
and discuss why it (as well as the kagom\'e lattice) behave differently
than the pyrochlore at anharmonic orders.
%% We compare the results to corresponding calculations on the two-dimensional
%% kagom\'e and checkerboard lattices and discuss the reasons that the pyrochlore
%% problem is harder to solve.
\end{abstract}

%75.25.+z       magnetic ordering
%75.10.Jm       Quantized spin models
%75.30.Ds       Spin waves (for spin-wave resonance, see 76.50.+g)
%75.40.Gb       Dynamic properties (dynamic susceptibility, spin waves, spin diffusion, dynamic scaling, etc.)
%75.50.Ee       antiferromagnetism

\maketitle

\section{\label{sec:intro}Introduction}

Highly frustrated magnetic systems 
are systems in which there is a zero temperature macroscopic classical
ground state degeneracy.\cite{phys_today,diep}
%%%%%%%%%%%%%%%%%%%%%%%%%%%%
\SAVE{Physics Today reference:
 revtex citations strip "month" or "number";  dirty trick put number in "volume" attribute.}
%%%%%%%%%%%%%%%%%%%%%%%%%%%%%
In experimental systems, this degeneracy is generically broken by secondary
interactions,
or by lattice distortions.\cite{motome_2005,tcher_distort,kotov}
However, even in toy models that include no such perturbations, one finds 
that the classical ground state degeneracy is broken by thermal fluctuations 
or quantum zero-point fluctuations.
Such phenomena are collectively referred to as
\emph{order by disorder}.\cite{villain,clh_obd}

Among three-dimensional geometrically frustrated systems, the most studied,
by far, is the pyrochlore lattice,
which is composed of the centers of the bonds of a 
diamond lattice,  so the pyrochlore sites form corner sharing tetrahedra.
%(see Fig.~\ref{fig:lattice}).
Despite numerous studies designed to illuminate on the ground state properties
of this model, in
the large-$S$ limit~\cite{moessner,clh_harmonic,uh_LN,uh_harmonic,motome_2005,
tcher_distort,capped_kagome,tcher_check},
a unique ground state has not been found for the pure,
undistorted, pyrochlore Heisenberg model.
In this paper, we answer this question by finding the
effective Hamiltonian that represents the 
quantum zero-point energy to {\it anharmonic} order
in spin waves.  (A short report has appeared in Ref.~\onlinecite{pyhfm}).

%\begin{figure} 
%\resizebox{7cm}{!}{\includegraphics{pyrochlore_gray.eps}}
%\caption{\label{fig:lattice}
%\footnotesize  The pyrochlore lattice}
%\end{figure}

We consider the nearest neighbor Heisenberg Hamiltonian on the pyrochlore
lattice
\begin{equation} \label{eq:heis}
\ham = J \sum_{\langle ij \rangle} \Svec_i \cdot \Svec_j \,.
\end{equation}
Here and below, $\langle ij \rangle$ denotes a sum over nearest neighbors.
Classically, all states satisfying
\begin{equation} \label{eq:classical_gs}
\sum_{i \in \alpha} \Svec_i=0\,,
\end{equation}
for all tetrahedra $\alpha$ 
are degenerate ground states,
with energy $-JN_s S^2$, where $N_s$ is the number of spins
(we reserve Greek indices for tetrahedra 
-- diamond lattice sites--
and roman indices for pyrochlore sites).

\subsection{Prior work} \label{sec:prior}

In recent work,\cite{clh_harmonic,uh_harmonic} we have studied the quantum
zero-point fluctuations of the large-$S$ limit of this model,
and found that, to harmonic order in the $1/S$ expansion, 
there remains an infinite degeneracy
of \emph{collinear} spin states (although the entropy of this family 
is non-extensive).
The degeneracy is associated with an exact invariance of the harmonic order 
energy to a gaugelike transformation.
Collinear configurations that are related by this symmetry have
identical \emph{fluxes} through all diamond lattice loops,
where the flux $\varphi_\LL$ through loop $\LL$ with bond centers at
$(i_1,i_2,\ldots,i_{2n})$ is defined as
\begin{equation} \label{eq:flux}
\varphi_\LL = \eta_{i_1} \eta_{i_2} \eta_{i_3} \cdots \eta_{i_{2n}} \,.
\end{equation}
The Ising variables $\eta_i \!=\! \pm 1$ correspond to the classical spin
direction along the collinearity axis.
The \emph{harmonic ground states} are all of the Ising configurations
in one of these
\emph{gauge families} and we call them the \emph{$\pi$-flux states},
following Ref.~\onlinecite{capped_kagome}.
These consist of all collinear configurations whose fluxes through all
\emph{hexagons} (the shortest diamond-lattice loops) are negative:
\begin{equation} \label{eq:pi_flux}
\prod_{i \in \hexagon} \eta_i =-1 \,, \qquad \forall \hexagon\,.
\end{equation}
[The argument for (\ref{eq:pi_flux}) is given in Sec.~\ref{sec:loop_expand}.]
Some of these states are shown in Fig.~9 of Ref.~\onlinecite{uh_harmonic}.
Furthermore, in Ref.~\onlinecite{uh_harmonic}, we constructed
an effective Hamiltonian for
the harmonic order zero-point energy, of the form
\begin{equation} \label{eq:heff_harm}
E_\harm^\eff = N_s S \left( E_0 + K_{6} \Phi_{6} + K_{8} \Phi_{8}
+\cdots \, \right)\,,
\end{equation}
where $E_0$, $K_n$ are numerical coefficients that can be evaluated
analytically~\cite{uh_harmonic}
($E_0\!=\!-0.5640$ , $K_6\!=\!0.0136 $, $K_8\!=\!-0.0033 $);
here $\Phi_{2n}$ is the total flux (per lattice site) through 
all diamond-lattice loops of length $2n$:
\begin{equation} \label{eq:Phi}
\Phi_{2n} \equiv
\frac{1}{N_s} \sum_{|\LL|=2n}
\varphi_\LL\,.
\end{equation}

In the interest of conciseness, throughout the rest of this paper
we use the term \emph{state} to mean ``classical Ising configuration''.
In this paper, we go beyond the harmonic order in the expansion $1/S$, to
search for a unique semiclassical ground state, focusing in the asymptotic 
$S \!\to\! \infty$ properties.
We consider small quantum fluctuations about classical Ising configurations
such that the local collinear order is preserved.
Our approach is aimed at deriving an
\emph{effective Hamiltonian}~\cite{clh_heff}
in terms of a much small number of degrees of freedom.

Similar work has been previously done on the closely related kagom\'e lattice.
This is a two-dimensional lattice, which is composed of corner sharing
triangles.
In the kagom\'e Heisenberg  antiferromagnet the zero-temperature classical
ground states satisfy~(\ref{eq:classical_gs}) for all \emph{triangles}
$\alpha$, and harmonic order spin-wave fluctuations select
\emph{all} coplanar classical configurations as degenerate ground states.
A self-consistent anharmonic theory breaks this degeneracy and selects one
unique coplanar ground state --the so-called $\sqrt{3}\!\times\!\sqrt{3}$
state.~\cite{chubukov,chan,chan_thesis}

\subsection{Outline of the paper}

This paper is organized as follows: in Sec.~\ref{sec:setup}
we derive the large-$S$ expansion by means of a Holstein-Primakoff
transformation. We review some of the results of
Ref.~\onlinecite{uh_harmonic}
on the harmonic theory.
In Sec.~\ref{sec:mft}, we derive the
mean-field Hamiltonian for the anharmonic theory, and present a self-consistent
variational scheme for solving it.

Then, in Sec.~\ref{sec:checker} we use a simple tractable example -- the
$(\pi,\pi)$ state on the two-dimensional checkerboard lattice--
in order to gain some analytic intuition on the behavior
of the two-point correlation functions that governs the mean-field 
quartic energy, and the scaling laws involved.
We find that these diverge as $\ln{S}$,
resulting in anharmonic energy of order $(\ln{S})^2$.
In Sec.~\ref{sec:checker_select} we argue that among all of the checkerboard 
lattice harmonic ground states, 
the quartic energy is minimized in the $(\pi,\pi)$ state, and
show numerical results to support this claim.
We find that, due to the different symmetries of the checkerboard lattice and the
Hamiltonian, the harmonic degeneracy in the checkerboard can be broken 
at the single-tetrahedron level, a result that cannot be carried over to the 
pyrochlore case.

In Sec.~\ref{sec:num} we present the main results of this paper.
--numerical calculations for the pyrochlore lattice.
We find that, as in the checkerboard, the quartic energy scales as
$(\ln{S})^2$. 
We calculate the anharmonic energy for a large set of harmonic ground states
and find that
and that the anharmonic theory breaks the degeneracy between them.
We derive effective Hamiltonians for both the gauge-invariant and
gauge-dependent terms in the quartic energy, and find a set of seemingly 
degenerate ground-states.

Next, in Sec.~\ref{sec:loop_expand}, we present a real-space loop expansion to
explain the nature of the dominant term in the gauge-dependent effective
Hamiltonian.
We analytically derive an effective Hamiltanion, which is different from the one we conjectured
in the numerical fitting.
Neverthless the leadng order terms of both effective Hamiltonians are minimized by the
same set of states which, as far as we can
tell, are all degenerate (both numerically and also
to very high order in the effective Hamiltonian).

\section{\label{sec:setup}Spin-wave theory}

In this section, we expand the Hamiltonian~(\ref{eq:heis}) in 
the semiclassical limit, in powers of $1/S$.
In Secs.~\ref{sec:harmonic} %and~\ref{sec:divergent}
we review some of the result in the
harmonic theory of Ref.~\onlinecite{uh_harmonic}, relevant to this paper.

\subsection{\label{sec:largeS}Large-$S$ expansion}

To study the quantum Heisenberg model, in the semiclassical limit of large $S$,
we perform the Holstein Primakoff transformation.
Since the harmonic ground states are all collinear,\cite{uh_harmonic}
we shall in the following, 
limit ourselves to states in which each site is labeled by an Ising variable
 $\eta_i$, such that, without loss of generality, the classical spin is
 $\Svec_i \!=\! \eta_i \hat{z}$, and
$\sum_{i \in \alpha} \eta_i\!=\!0$ for any tetrahedron $\alpha$.
Thus each tetrahedron includes four satisfied -- antiferromagnetic
(AFM) -- bonds
and two unsatisfied -- ferromagnetic (FM) -- bonds.
Notice that, whenever the spins satisfy the classical
ground state condition (\ref{eq:classical_gs}),
the sum of neighbor spins is $(-2)$
times the spin on a site, i.e.
 \begin{equation} \label{eq:nbrsum}
  \sum_{j {\rm n.n.~of} i} \eta_j = - 2\eta_i .
 \end{equation}

We first rotate the local coordinates to 
$(\eta_i \hat{x},\hat{y},\eta_i \hat{z})$,
 and define boson operators $a_i$, $\adag_i$ such that
\begin{eqnarray}
S^z_i &=& \eta_i (S- \adag_i a_i) \,, \nonumber \\
S^+_i &\equiv& \eta_i S^x + i S^y =
  \sqrt{2S-\adag_i a_i} \,\, a_i 
 \,, \nonumber \\
S^-_i &\equiv& \eta_i S^x - i S^y = \adag_i \sqrt{2S-\adag_i a_i} \,. 
\label{eq:holstein}
 \end{eqnarray}
These operators satisfy the canonic bosonic commutation relations
\begin{equation}
[a_i,\adag_j] =\delta_{ij} \,, \qquad 
[a_i,a_j] \!=\!0 \,, \qquad
[\adag_i,\adag_j] \!=\!0 \,.
\end{equation}
We now expand Eq.~(\ref{eq:holstein}) in powers of $1/S$,
and express the Hamiltonian in terms of spin deviation operators 
\begin{equation}
\sigma^x_i = \eta_i \sqrt{\frac{S}{2}} (a_i+\adag_i) \,, \qquad
 \sigma^y_i = -i \sqrt{\frac{S}{2}} (a_i-\adag_i) \,, \label{eq:sdev}
\end{equation}
and obtain~\cite{kvale}
\begin{subequations}
\label{eq:ham_expand}
\begin{eqnarray}
\label{eq:ham0}
\ham&=&E_\cl+ \ham_\harm + \ham_\quart +O(S^{-1}) \,, \\
E_\cl&=& -J N_s S^2 \,,  \\
\label{eq:ham2}
\ham_\harm &=& \tJ  \sum_i ((\sigma^x_i)^2 + (\sigma^y_i)^2)
 + \tJ  \sum_{\langle ij \rangle} 
 (\sigma^x_i \sigma^x_j + \sigma^y_i \sigma^y_j ) \nonumber \\
  &&-\tJ S N_s \,, \\
\ham_\quart &=& 
 \frac{\tJ}{8 S^2} \sum_{\langle ij\rangle} \Big[ 2 \eta_i \eta_j
\big((\sigma^x_i)^2 + (\sigma^y_i)^2\big) 
\big((\sigma^x_j)^2 + (\sigma^y_j)^2\big) 
 \nonumber \\
 &-& \!\!
 \sigma^x_i \big((\sigma^x_j)^3\!+\! \sigma^y_j \sigma^x_j \sigma^y_j\big) -\!
 \sigma^x_j \big((\sigma^x_i)^3\!+\! \sigma^y_i \sigma^x_i \sigma^y_i\big) 
\nonumber \\
 &-& \!\! 
 \sigma^y_i \big((\sigma^y_j)^3\!+\! \sigma^x_j \sigma^y_j \sigma^x_j\big) -\!
 \sigma^y_j \big((\sigma^y_i)^3\!+\! \sigma^x_i \sigma^y_i \sigma^x_i\big)  
   \Big] \,. 
\label{eq:ham4}
\end{eqnarray}
\end{subequations}
where 
$\tJ\!\equiv\!J(1\!+\!1/2S)$.
In the following, we shall set $\tJ\!=\!1$.
Somewhat redundantly, we also define $\tJ_{ij} \!\equiv \! \tJ \!=\! 1$ when 
$(i,j)$ are nearest neighbors, and zero otherwise (to simplify sums
over just one index.)

\subsection{\label{sec:harmonic}Harmonic Hamiltonian}
The use of the operators $\sigma^x$, $\sigma^y$ allows us to represent
the harmonic Hamiltonian~(\ref{eq:ham2}) in block diagonal form 
\begin{equation}
\ham_\harm= 
 \left(  (\svec^x)^\dagger , (\svec^y)^\dagger  \right)
\left( \begin{array}{cc}
\Hvec & 0 \\ 0 & \Hvec \end{array} \right)
 \left( \begin{array}{c} \svec^x \\ \svec^y \end{array} \right)
- \tr{\Hvec} \,,
\label{eq:harm_mat}
\end{equation}
where $\svec^x$ and $\svec^y$ are vector operators with respect to
site indices, of length $N_s$,
%
%\mbox{ and} \qquad
and the $N_s\!\times\!N_s$ matrix $\Hvec$ has elements
\begin{equation}
\label{eq:heis_mat}
H_{ij} = \left\{ \begin{array}{ll}
1 & i=j \\
\frac{1}{2} & i,j \mbox{ nearest neighbors} \\
0& \mbox{otherwise} \end{array} \right. \,.
\end{equation}
The dependence on the particular classical ground
state comes via the commutation relations
\begin{equation} \label{eq:sig_comm}
[\sigma^x_i,\sigma^y_j] = i S \eta_i \delta_{ij} \,.
\end{equation}
In Ref.~\onlinecite{uh_harmonic} we detailed the harmonic theory and the
properties of the eigenmodes.
Here we briefly summarize the results relevant to this paper, for completeness.

\subsubsection{Diagonalization} \label{sec:diag}

Define the $N_s\!\times\!N_s$
diagonal matrix $\etvec$ by 
$\eta_{ij}\!\equiv\!\eta_i \delta_{ij}$.
Then spin-wave modes of any Hamiltonian of the form~(\ref{eq:harm_mat}),
with operator commutation relations~(\ref{eq:sig_comm}) are the
eigenvectors $\{\vvec_m\}$, with eigenvalues $\{\lambda_m\}$, 
of the {\it dynamical matrix} $\etvec \Hvec$:
\begin{equation} \label{eq:eig}
\eta_i \lambda_m v_m(i) =v_m(i) + \half \sum_j \tJ_{ij} v_m(j)\,.
\end{equation}
The eigenvectors satisfy a pseudo orthogonality constraint
\begin{equation} \label{eq:orth}
\vvec_l^\dagger \etvec \vvec_m \propto \delta_{lm}\,.
\end{equation}
The corresponding frequencies are $\hbar \omega_m\!=\! 2S|\lambda_m|$,
and thus the zero-point energy is
\begin{equation} \label{eq:zero_point}
E_\harm=S \sum_m \Big(|\lambda_m|-1\Big) \,.
\end{equation}
In Refs.~\onlinecite{uh_harmonic} and \onlinecite{clh_harmonic},
it was shown that the zero point 
energy is minimized for configurations that satisfy~(\ref{eq:pi_flux}). 
A condensed version of this derivation shall be given later,
in Sec.~\ref{sec:loop_harm}.

For the Heisenberg Hamiltonian matrix~(\ref{eq:heis_mat}) on the 
pyrochlore lattice, one finds that 
(for \emph{any} Ising ground state)
half the spin-wave modes have vanishing frequencies.
These are the \emph{zero modes}, which satisfy
\begin{equation}\label{eq:zeromode}
\sum_{i \in \alpha} v_m(i) = 0\,.
\end{equation}
for all tetrahedra $\alpha$.
The two-point correlations (fluctuations)
$G_{ij}$ of the spin deviation operators, it can be shown, are
given by 
\begin{eqnarray}
\mathbf{G}\equiv \langle \svec^x (\svec^x)^\dagger \rangle =
\langle \svec^y (\svec^y)^\dagger \rangle
&=& \sum_m
\frac{S}{2}\frac{\vvec_m \vvec_m^\dagger}{|\vvec_m^\dagger \etvec \vvec_m|}
 \,, \nonumber \\
\langle \svec^x (\svec^y)^\dagger +  \svec^y (\svec^x)^\dagger \rangle
&=& \Ovec \,.
\label{eq:fluct}
\end{eqnarray}
It is clear from~(\ref{eq:fluct}) that any mode $\vvec_m$ for
which $\vvec_m^\dagger \etvec \vvec_m\!=\!0$, exhibits
divergent fluctuations.
We call such a mode a \emph{divergent mode} and it turns out that such a mode
is necessarily a zero mode, i.e. $\lambda_m\!=\!0$.
The converse is not true-- most zero modes have nonsingular fluctuations.

\subsubsection{Ordinary modes}\label{sec:ordinary}

The eigenmodes of Eq.~(\ref{eq:eig}) can be divided into two groups:
half ($N_s/2$) of the modes have zero frequency. 
We call these \emph{generic zero modes},\cite{uh_harmonic} because
the subspace that they span 
is identical for any collinear classical ground
state.~\footnote{
%%%%%%%%%%%%%%%%%%%%%%%%
Note that the choice of a \emph{basis} within the subspace of
generic zero modes \emph{does} depend on the particular collinear state, as the 
pseudo orthogonality condition~(\ref{eq:orth}) depends on $\etvec$.}
%%%%%%%%%%%%%%%%%%%%%%%%
Since these modes have zero frequency, they do not contribute to the harmonic
zero-point energy.

The other half of the modes are called \emph{ordinary modes},\cite{uh_harmonic}
and these modes can be naturally expressed in terms of 
\emph{diamond-lattice} modes (recall that the diamond lattice has
$N_s/2$ sites):
an (un-normalized) ordinary mode $\vvec_m$ can be written down as 
\begin{equation} \label{eq:ord_modes}
v_m(i) = \frac{1}{\sqrt{2}} \eta_i \sum _{\alpha: i\in \alpha}
u_m(\alpha)
\end{equation}

%
%%% \begin{equation} \label{eq:ord_modes}
%%% v_m(i) = \frac{1}{\sqrt{2}} \eta_i [u_m(\alpha(i))+u_m(\beta(i))] \,,
%%% \end{equation}
%
where the sum runs over the two tetrahedra to which site $i$ 
belongs and
$\uvec_m$ is a vector of length $N_s/2$, living on the centers
of tetrahedra (diamond lattice sites), and satisfying the spin-wave equation
\begin{equation}\label{eq:sw_diamond}
\lambda_m u_m(\alpha) = \half {\sum_\beta}' \eta_{i(\alpha\beta)}
u_m(\beta) \,,
\end{equation}
where the sum is over (diamond-lattice) nearest neighbors of $\alpha$, and
$i(\alpha\beta)$ is the pyrochlore site on the center of the bond 
connecting $\alpha$ and $\beta$.
The diamond-lattice modes $\{\uvec_m\}$ are eigenmodes of an Hermitian matrix
and therefore are orthogonal to each other in the usual sense.
We choose the normalization  $|\uvec_m|=1$ without loss of generality.
From ~(\ref{eq:ord_modes}) and (\ref{eq:sw_diamond}) one easily simplifies
the pseudo-norm denominator in Eq.~(\ref{eq:fluct}),
\begin{equation}\label{eq:eta-norm-ord}
 \vvec_m^\dagger \etvec \vvec_m= \lambda_m
\end{equation}
(valid only for ordinary modes).

It is evident that the solutions of Eq.~(\ref{eq:sw_diamond}) are
invariant under a \emph{gaugelike transformation} of the state:
if we transform $\eta_i \to \tau_\alpha \tau_\beta \eta_i$, where 
$\tau_\alpha,\tau_\beta = \pm 1$, then the dispersion would not change,
and the ordinary modes would transform
$u_m(\alpha) \to \tau_\alpha u_m(\alpha)$.
Taking $\tau_\alpha=-1$ amount to flipping all of the spins in tetrahedron 
$\alpha$. 
Such a transformation is not literally a gauge transformation
since the flips must be correlated, so that the tetrahedron rule 
-- $\sum _{i\in\alpha} \eta_i=0$ from (\ref{eq:classical_gs}) --
is preserved.
Whenever two states are related by a gaugelike transformation, 
they have the same spin-wave eigenvalues $\lambda_m$ and hence
identical values of the total harmonic zero-point energy.

Although most of the ordinary modes carry nonzero frequency, there is a subset
of them that has $\lambda_m=0$. It turns out that these are the divergent modes
-- modes that have $\vvec_m^\dagger \etvec \vvec_m=0$, and whose
correlations are divergent [see Eqs.~(\ref{eq:fluct}) 
and (\ref{eq:eta-norm-ord})].

\subsubsection{Fourier transformed Hamiltonian}
In order to perform numerical calculations on large systems, 
we must limit ourselves to periodic states.
We shall assume a \emph{magnetic unit cell} with $N_M$ sites
arranged on a \emph{magnetic lattice}.
In the simplest possible, $\Qvec\!=\!\Ovec$ case, $N_M\!=\!4$ and the magnetic
lattice is the fcc.
Most of this work focuses on harmonic ground states, i.e.,
$\pi$-flux states. The \emph{smallest} possible unit cell for that case has
$N_M\!=\!16$ sites.
In practice, the calculation can often be simplified by utilizing the
\emph{bond order}, which may have a smaller unit cell.~\cite{uh_thesis}

We Fourier transform the Hamiltonian~(\ref{eq:harm_mat}):
\begin{eqnarray} \label{eq:harm_mat_q}
\ham_\harm  &=&
\sum_\qvec
\left( (\svec^x_\qvec)^\dagger,(\svec^y_\qvec)^\dagger \right)
\left( \begin{array}{cc}
\Hvec(\qvec) & \Ovec \\
\Ovec & \Hvec(\qvec) 
\end{array} \right)
\left( \begin{array}{c}
\svec_{\!-\qvec}^x \\ 
\svec_{\!-\qvec}^x \end{array} \right) \nonumber \\
&&-\tr{\Hvec(\qvec)} \,, 
\end{eqnarray}
where $\svec_\qvec^x$, $\svec_\qvec^y$ are vectors of length $N_M$
of the Fourier transformed $x$ and $y$ spin deviation operators.
The wavevector $\qvec$ is in the Brillouin zone of the magnetic lattice.
\begin{eqnarray}
\vec{\sigma}_i
&=& \frac{1}{\sqrt{N_M}} \sum_{\qvec} \vec{\sigma}^{l_i}_{\qvec}
e^{-i \qvec \cdot [\Rvec_i + \Dvec_{l_i})}\,,\nonumber \\
\vec{\sigma}^l_{\qvec} &=& \frac{1}{\sqrt{N_M}}
\sum_{\Rvec} \vec{\sigma}^l_{\Rvec}
 e^{i \qvec \cdot (\Rvec + \Dvec_l)}\,,
 \label{eq:fourier_sigma}
\end{eqnarray}
where $\Rvec$ is a magnetic lattice vector and $l$ is a sublattice index,
corresponding to a basis vector $\Dvec_l$, 
i.e., for site $i$: $\rvec_i\!=\!\Rvec_i\!+\!\Dvec_{l_i}$

Upon diagonalization of the Hamiltonian [i.e., finding eigenmodes
of $\etvec \Hvec(\qvec)$, where $\etvec$ is now $N_M\!\times\!N_M$],
we obtain $N_M$ bands within the Brillouin zone,
half of which are of zero mode bands, and half are of ordinary modes.
%We refer to modes in the other, non-zero bands as \emph{ordinary modes}.
The divergent spin-wave modes can be shown to occur along \emph{lines} in the
Brillouin zone where an ordinary mode frequency goes to
zero (we call these \emph{divergence lines}).~\cite{uh_harmonic}
Each of these divergence lines is parallel to one of $x$, $y$, or $z$ axes.

The correlations of spin fluctuations can be expanded in terms of Fourier
components, using Eq.~(\ref{eq:fourier_sigma}):
\begin{equation}
G_{ij} \equiv \langle \sigma_i \sigma_j \rangle
\frac{N_M}{N_s}\sum_\qvec G_{l_i l_j}(\qvec) \cos \chivec_{ij} \cdot \qvec \,,
\end{equation}
with
\begin{equation} \label{eq:corrq}
G_{l_i l_j}(\qvec)\equiv
\langle \sigma_\qvec (l_i) \sigma_{-\qvec}(l_j)
\rangle \,,
\end{equation}
where $l_i$ and $l_j$ are the sublattice indices of $i$ and $j$, respectively,
and $\chivec_{ij}\!=\!\rvec_i\!-\!\rvec_j$.

\section{Self-consistent anharmonic theory} \label{sec:mft}

This section develops our mean-field prescription 
to self-consistently 
calculate the anharmonic corrections to the energy, for an 
arbitrary given state $\{\eta_i\}$.
First, (Sec.~\ref{sec:mf}) we decouple the quartic term $\ham_\quart$ and 
write down a quadratic mean-field Hamiltonian. 
Next, we introduce a variational Hamiltonian as an approximation
for mean-field problem (Sec.~\ref{sec:ham-var}), and
in Sec.~\ref{sec:selfcons-OK} show that the variational form
agrees with a general self-consistent approach in the
large-$S$ limit.
In Sec.~\ref{sec:scaling} we discuss how various fluctuations and
energy scales depend on $S$.

\subsection{Decoupling  scheme}

First let us work through the Hartree-Fock-like decoupling of the 
quartic term~(\ref{eq:ham4}) of our spin-wave Hamiltonian~\cite{uh_thesis}.
It turns out the decoupled coefficients depend on the (Ising)
spin configuration in a simple fashion (Sec.~\ref{sec:Gamma-form})
which allows us (in principle) to reduce the self-consistency 
conditions to a one-parameter equation.

\subsubsection{Energy expectation and decoupled Hamiltonian}\label{sec:mf}

In a decoupling, one implicitly assumes a variational 
wavefunction $\Psi_\MF$, a priori unconstrained except for being Gaussian.
Thus, it is specified by a harmonic effective Hamiltonian $\ham_\MF$,
defined so that 
\begin{equation} \label{eq:decoupling_condition}
\langle \ham_\harm + \ham _\quart \rangle = \langle \ham _\MF \rangle 
\end{equation}
where the expectations are taken with respect to $\ham_\MF$ itself.

In light of Wick's theorem, we can immediately write the 
energy expectation by plugging 
into (\ref{eq:ham2}) and (\ref{eq:ham4})
the two-point correlations defined in (\ref{eq:fluct}), 
but now using the $\ham_\MF$ wavefunction:
\begin{subequations}
\label{eq:hamexp}
\begin{eqnarray}
%% \label{eq:hamexp_harm}
\langle \ham_\harm \rangle &=&
  2 \Bigl(\sum _i  G_{ii} + \sum _{\langle ij\rangle} G_{ij}
  - S N_s \Bigr) \\
\langle \ham_\quart \rangle &=&
    \frac{1}{2S^2}  \sum _{\langle ij \rangle} 
    \big[\eta_i \eta_j (G_{ii}G_{jj} + G_{ij}^2) 
           - G_{ij}(G_{ii}+G_{jj})\big]\nonumber\\
%% \label{eq:hamexp_quart}
\end{eqnarray}
\end{subequations}
To make some expressions more compact, 
we define a bond variable, 
\begin{equation} \label{eq:corrdiff}
\Gamma_{ij} \equiv G_{ii}-\eta_i \eta_j G_{ij}\,.
\end{equation}
$\Gamma_{ij}$ is, in general, {\it not} symmetric~\footnote{
%%%%%%%%%%
$\Gamma_{ij}$ becomes symmetric in the large-$S$/small-$\varepsilon$ limit
we are interested in: see Eqs.~(\ref{eq:gamma_emp}) and (\ref{eq:gauge_emp}).}
%%%%%%%%%%%%%%
and is defined only for $(i,j)$ nearest neighbors (nonzero $\tJ_{ij}$).
%%% Notice that, whenever the $\{\eta_i \}$ satisfy the classical
%%% ground state condition, the sum of $\eta_j$ over neighbors of
%%% $\eta_i$ gives $-2\eta_i$, 

Substituting (\ref{eq:corrdiff}) into (\ref{eq:hamexp}),
and using (\ref{eq:nbrsum}), we get
\begin{subequations}
\label{eq:hamexp_Gamma}
\begin{eqnarray}
%% \label{eq:hharmexp_Gamma}
\langle \ham_\harm \rangle &=& 
- \sum_{\langle ij\rangle} 
   \Big[\eta_i \eta_j \left( \Gamma_{ij}+\Gamma_{ji}\right)-
    SN_s\Big];\\
\langle \ham_\quart \rangle &=& 
\frac{1}{S^2}
\sum_{\langle ij\rangle} \eta_i \eta_j \Gamma_{ij}\Gamma_{ji}.
\label{eq:hquartexp_Gamma}
\end{eqnarray}
\end{subequations}

Then 
\begin{eqnarray} \label{eq:emf}
E_\MF&\equiv& \langle \ham_\MF \rangle =
-\sum_{ij} (H_\MF)_{ij} G_{ij} =
\\ &&
-\sum_{\langle ij \rangle} \eta_i \eta_j
\left( \Gamma_{ij} + \Gamma_{ji} - \frac{1}{S^2} \Gamma_{ij} \Gamma_{ji} \right) -S N_s;
\,.\nonumber
\end{eqnarray}
and [using~(\ref{eq:hamexp})]
we see indeed $\ham_\MF$ satisfies (\ref{eq:decoupling_condition}).

To write our decoupled Hamiltonian $\ham_\quart+\ham_\harm$, 
we adopt a matrix form, in
analogy with the harmonic Hamiltonian~(\ref{eq:harm_mat})
\begin{equation}
\ham_\MF= 
 \left(  (\svec^x)^\dagger , (\svec^y)^\dagger  \right)
\left( \begin{array}{cc}
\Hvec_\MF & \Ovec \\
\Ovec & \Hvec_\MF \end{array} \right)
 \left( \begin{array}{c} \svec^x \\ \svec^y \end{array} \right)
- SN_s \,;
\label{eq:hmf}
\end{equation}
defining the matrix elements in Eq.~(\ref{eq:hmf}) to depend on the 
correlations $G_{ij}$:
\begin{subequations}
\label{eq:hmf_elts}
\begin{eqnarray}
\label{eq:hmf_ij}
(H_\MF)_{ij} &=& \frac{\tJ_{ij}}{2}\left[1-
\frac{G_{ii}+G_{jj} -2 \eta_i \eta_j G_{ij}}{2S^2}
\right]\,,\\
\label{eq:hmf_ii}
(H_\MF)_{ii} &=& 1+\frac{1}{2S^2} \sum_j \tJ_{ij}
\left(\eta_i \eta_j G_{jj} - G_{ij} \right) \,.
\end{eqnarray}
\end{subequations}
Recall from Sec.~\ref{sec:largeS} that
$\tJ_{ij}=1$ for nearest neighbors, otherwise zero.
Thus, although $G_{ij}$ decays as a power law,
$\ham_\MF$ has only on-site and nearest-neighbor terms.
In terms of the $\Gamma_{ij}$ variables, eq.~(\ref{eq:hmf_elts})
reads
%%%%%%%%%%%%%%
\begin{subequations}
\label{eq:hmf_gamma}
\begin{eqnarray}
\label{eq:hmf_gamma_ij}
(H_\MF)_{ij} &=& \frac{\tJ_{ij}}{2} \left[1 - \frac{1}{2S^2}
(\Gamma_{ij}+\Gamma_{ji})\right]  \\
(H_\MF)_{ii} &=& 1+ \frac{1}{2S^2} \sum_j \tJ_{ij} \eta_i \eta_j \Gamma_{ji}
\label{eq:hmf_gamma_ii}
\end{eqnarray}
\end{subequations}
%%%%%%%%%%%%%%%%%%%%%%%%

All the machinery that was applied to $\Hvec$ for the harmonic problem 
in Sec.~\ref{sec:harmonic}, can now be applied to $\Hvec_\MF$.
In particular, we can evaluate the correlations $\{ G_{ij} \}$,
in terms of which the Hamiltonian matrix elements are written.
Thus, by the self-consistent decoupling approximation we have 
replaced the interacting spin-wave Hamiltonian by an 
effective non-interacting theory.

Unfortunately, this does not yet give a solution,
since the $\{G_{ij}\}$ are {\it a priori} unknown. 
We cannot just use the correlations obtained from the bare harmonic
theory~(\ref{eq:harm_mat}) for both practical reasons ($G_{ij}$ diverges
in that case)
and substantive ones: the theory would not be self-consistent -- we would not
recover the same correlations as those we put into it.
A solution may, in fact be obtained by successive iterations:
assume a trial set of coefficients $\Hvec_\MF$, compute the implied
correlations, and define the next iteration of $\Hvec_\MF$ 
from (\ref{eq:hmf_ij}).

\subsubsection{Simplified form of ~$\Gamma_{ij}$ and $\ham_\MF$} 
\label{sec:Gamma-form}

In principle this iteration seems forbidding, but it is 
simplified by an important fact, discovered numerically
but verified analytically.  For any $\ham_\MF$ approaching
$\ham_\harm$, as should be the case for large $S$:
\begin{equation} \label{eq:gamma_emp}
 \Gamma_{ij} = \Gamma^{(0)} +
 \Gamma^{(2)}  \eta_i \eta_j + \Delta \Gamma_{ij}
%  \Gamma_{ij} = \Gamma^{(0)} (\varepsilon)+
%  \Gamma^{(2)} (\varepsilon) \eta_i \eta_j + \Delta \Gamma_{ij}(\varepsilon)
\,.
\end{equation}
Here $\Gamma^{(0)}$ and $\Gamma^{(2)}$ are diverging terms
independent of $i$, $j$ (and of the same order);
whereas $\Delta \Gamma_{ij}$ does depend on $i$ and $j$.
but is much smaller than $\Gamma^{(2)}$.
This was seen numerically in the outputs from a particular family
of starting parameters, the family of variational wavefunctions
$\Psi(\varepsilon)$ specified by $\ham_\var (\varepsilon)$ 
[defined below in Sec.~\ref{sec:ham-var}].
More generally, an analytic explanation of the form~(\ref{eq:gamma_emp}), 
i.e.  why $\Gamma_{ij}$ depends only on $\eta_i \eta_j$ at
dominant order, is found in Appendix~\ref{app:ordinary}.
[It follows from the gaugelike  invariance,
for the special case of Ising configurations that 
minimize the harmonic energy, the $\pi$-flux states.
One might crudely paraphrase that argument 
by saying the correlations that come out of the bare Hamiltonian
have the form~(\ref{eq:gamma_emp}) (albeit with divergent
$\Gamma^{(0)}$, $\Gamma^{(2)}$).

Next, inserting the relation~(\ref{eq:gamma_emp}) into
%%% Eqs.~(\ref{eq:hmf_ii}) and~(\ref{eq:hmf_ij}),
Eqs.~(\ref{eq:hmf_gamma}), 
we can write the matrix elements of the mean-field Hamiltonian
%
%%% \begin{subequations}
%%% \begin{eqnarray}
%%% &&(H_\MF)_{ij} = \frac{\tJ_{ij}}{2} \left[1 - \frac{1}{2S^2}
%%% \label{eq:hmf_gamma12_ij}
%%% (\Gamma_{ij}+\Gamma_{ji})\right]  \\ &=&
%%% \frac{\tJ_{ij}}{2} \left[1- \frac{1}{S^2} \Gamma^{(0)}
%%% - \frac{1}{S^2} \Gamma^{(2)}  \eta_i \eta_j -
%%% \frac{1}{2S^2}  (\Delta \Gamma_{ij} + \Delta \Gamma_{ji})\right]  \nonumber\\
%%% &&(H_\MF)_{ii} = 1+ \frac{1}{2S^2} \sum_j \tJ_{ij} \eta_i \eta_j \Gamma_{ji}
 %%% \\ \nonumber &=&
%%% 1 - \frac{1}{S^2}\Gamma^{(0)} +
%%% \frac{3}{S^2} \Gamma^{(2)}
%%% + \frac{1}{2S^2} \sum_j \tJ_{ij} \eta_i \eta_j \Delta \Gamma_{ji} \,.
%%% \label{eq:hmf_gamma12_ii}
%%% \end{eqnarray}
%%% \end{subequations}
\begin{subequations}
\label{eq:hmf_gamma12}
\begin{eqnarray}
\label{eq:hmf_gamma12_ij}
(H_\MF)_{ij} &=&
\frac{\tJ_{ij}}{2} \Bigg[ \Big(1- \frac{1}{S^2} \Gamma^{(0)} \Big)
- \frac{1}{S^2} \Gamma^{(2)}  \eta_i \eta_j  \\
&-& \frac{1}{2S^2}  \big(\Delta \Gamma_{ij} + \Delta \Gamma_{ji}\big)\Bigg] \nonumber  \\
(H_\MF)_{ii} &=& \Big( 1 - \frac{1}{S^2}\Gamma^{(0)}\Big) +
\frac{3}{S^2} \Gamma^{(2)} \nonumber\\
&+& \frac{1}{2S^2} \sum_j \tJ_{ij} \eta_i \eta_j \Delta \Gamma_{ji} \,.
\label{eq:hmf_gamma12_ii}
\end{eqnarray}
\end{subequations}
To get the last line of Eq.~(\ref{eq:hmf_gamma12_ii}), 
we used the $z=6$ coordination of the
pyrochlore lattice, and the classical tetrahedron constraint
$\sum _{i\in\alpha} \eta_i=0$ [from (\ref{eq:classical_gs})].
We now define
\begin{equation}
J^* \equiv 1-\frac{1}{S^2} \Gamma^{(0)} \,, \qquad J^*_{ij}\equiv J^* \tJ_{ij} \,.
\end{equation}
Note that $|J^*-1|\ll 1$.  We obtain
 \begin{subequations}
 \label{eq:hmfout_all}
 \begin{eqnarray}
 %%\label{eq:hmfout_ij}
 (H_\MF)_{ij} &=&
\frac{J^{*}_{ij}}{2} \left(1 - \eta_i \eta_j \right) \\ &&-
 \frac{1}{2S^2}\left(\Delta \Gamma_{ij} + \Delta \Gamma_{ji}\right) \nonumber \\
 %%\label{eq:hmfout_ii}
 (H_\MF)_{ii} &=&  J^* \left( 1+ \frac{3}{4} \varepsilon_\out \right)
  \\ \nonumber &&+
  \frac{1}{2S^2} \sum_j \tJ_{ij} \eta_i \eta_j \Delta \Gamma_{ji} \,.
 \end{eqnarray}
 \end{subequations}
%
%%% \begin{subequations}
%%% \label{eq:hmfout_all}
%%% \begin{eqnarray}
%%% %%\label{eq:hmfout_ij}
%%% (H_\MF)_{ij} &=&
%%% \frac{J^{*}_{ij}}{2} \left(1
%%% - \frac{1}{S^2 J^*}\Gamma^{(2)} \eta_i \eta_j \right) \\ &&-
%%% \frac{1}{2S^2}\left(\Delta \Gamma_{ij} + \Delta \Gamma_{ji}\right)  \nonumber\\
%%% %%\label{eq:hmfout_ii}
%%% (H_\MF)_{ii} &=&  J^* \left( 1+ \frac{3}{S^2 J^*}\Gamma^{(2)} \right)
 %%% \\ \nonumber &&+
 %%% \frac{1}{2S^2} \sum_j \tJ_{ij} \eta_i \eta_j \Delta \Gamma_{ji} \,.
%%% \end{eqnarray}
%%% \end{subequations}
where 
\begin{equation}
  \varepsilon_\out
\! \equiv \!   \frac{4 \Gamma^{(2)}}{ S^2 J^* } .
\end{equation}
%%% where the $\varepsilon$ dependence is shown explicitly.
Thus, if we drop the much smaller terms in $\Delta \Gamma_{ij}$
all the corrections are proportional to a single parameter $\Gamma^{(2)}$
times simple functions of the spin configuration.

\subsection{Variational Hamiltonian} \label{sec:ham-var}

The one-parameter dependence of 
Eq.~\eqr{eq:hmfout_all} suggests we do not need to explore 
the full parameter space of trial Hamiltonians to find
the self-consistent mean-field Hamiltonian.
Instead, we shall limit ourselves to a simplified 
variational Hamiltonian $\ham_\var$, which though it has just one variational parameter, 
appears to capture all the important properties of $\ham_\MF$.
(Specifically, $\ham_\var$ approximates $\ham_\MF$ better and better in the
limit $S\to \infty$, as will be shown analytically below.)

So, we wish to write a harmonic $\ham_\var$, as simple as possible,
to specify the Gaussian variational wavefunction $\Psi_\var$,
its ground state (not necessarily equal to $\Psi_\MF$).
Since $\ham_\MF$ -- the solution to an unconstrained
variational problem -- has only nearest-neighbor terms,
there is no loss of generality
when we restrict our variational search to that form.
[In contrasted,  on the kagom\'e lattice, 
the appropriate variational Hamiltonian had
second- or third-nearest-neighbor (Heisenberg)
terms~\cite{harris,chubukov,chan,chan_thesis},
due to cubic terms in the spin-wave expansion.]
We thus adopt the simplest nontrivial form, 
the same as~(\ref{eq:hmf}), except with the diagonal block matrix $\Hvec_\MF$
replaced by
\begin{equation}
\label{eq:variational}
\Hvec_\var \equiv \Hvec + \delta \etvec \Hvec \etvec
+ \varepsilon \openone\,. 
\end{equation}
where $\delta$ and $\varepsilon$ are variational parameters.
The $\delta$ modifies the strength of 
AFM and FM bonds  in opposite ways:
namely, $(H_\var)_{ij}\!=\!(1\!+\!\delta)/2$
for neighbors with $\eta_i=\eta_j$ and
$(H_\var)_{ij}\!=\!(1\!-\!\delta)/2$ for neighbors with $\eta_i=-\eta_j$.
This is the simplest possible form of a variational Hamiltonian that is
consistent with the local spin symmetries.

We do require invariance under global spin rotations,
which means the Goldstone mode (associated with
global rotation) must have zero energy.  
Its eigenvector $\vvec_G$ has elements
\begin{equation} \label{eq:goldstone}
v_G(i) = \frac{\eta_i}{\sqrt{N_s}}\,, \qquad \forall i \, .
\end{equation}
%
%%% $\etvec\Hvec_\var \vvec_G =0$.
Thus we require $\etvec\Hvec_\var \vvec_G =0$; inserting 
Eq.~(\ref{eq:hmf_ii}) and writing out each term, we first
note $\Hvec \vvec_G=0$ so our condition is
\begin{equation}
0=\eta_i \sum_j (H_\var)_{ij} v_G(j)
%\eta_i \sum_j \left( \eta_j H_{ij} +
%\delta \eta_i H_{ij}  + \varepsilon \delta_{ij} \eta_j \right )
= 4 \delta  +\varepsilon
\end{equation}
Thus (\ref{eq:variational})
ends up having only one independent variational parameter $\varepsilon$.
It will become clear in the following, that the correct signs 
for the parameters are $\varepsilon>0$, $\delta<0$.
So, just writing out the components of $\Hvec_\var$ as 
defined in (\ref{eq:variational}),
\begin{subequations}
\label{eq:hvar_all}
\begin{eqnarray}
\label{eq:hvar_ij}
(H_\var)_{ij} &=& \frac{1}{2}\left(1-\frac{\varepsilon}{4}\eta_i\eta_j\right)
\,,\\
\label{eq:hvar_ii}
(H_\var)_{ii} &=& 1+\frac{3}{4} \varepsilon \,.
\end{eqnarray}
\end{subequations}
Note that $(ij)$ in (\ref{eq:hvar_ij}), and in similar equation pairs,
applies only to nearest-neighbor sites.

A more elaborate (multi-parameter) trial form of $\ham_\var$
might improve the quality of the calculation, 
by exploring a larger set of variational wavefunctions;
this is particularly important when the 
Ising configuration is not uniform from the gauge-invariant
viewpoint (see Appendix~\ref{app:gauge-correl}), 
since \eqr{eq:gamma_emp} breaks down in that case.
Nevertheless, as we shall see numerically in Sec.~\ref{sec:num},
the most important degeneracy-breaking effects
are captured within this simple one-parameter theory.

\subsubsection{Self-consistent approach}

Revisiting eqs.~(\ref{eq:hmfout_all}), we see they reduce to 
Eqs.~(\ref{eq:hvar_all}) but with $\varepsilon \to \varepsilon_\out$.
Furthermore, as  $\varepsilon\to 0$, it turns out 
$\varepsilon_\out(\varepsilon)$ is {\it increasing}
[indeed logarithmically divergent: 
see ~(\ref{eq:gij_r_check}) and (\ref{eq:gij_r})].
%% for the cases of the checkerboard or pyrochlore lattices.]
So there is a unique self-consistent solution to
\begin{equation} \label{eq:vareps-selfcons}
\epsSC
= \varepsilon_\out(\epsSC) = 
\frac{4 \Gamma^{(2)}(\epsSC)}{ S^2 J^* } .
\end{equation}
and at $\varepsilon=\epsSC$, 
[neglecting the $\Delta \Gamma_{ij}$) correction  terms]
we get
\begin{equation} \label{eq:HMF-agrees}
\Hvec_\MF \approx J^* \Hvec_\var .
\end{equation}
%%%%%%%%%%%%%%%%%%%%%%%%
Of course, the overall prefactor of $J^*$ has no effect 
on the spin correlations comprising $\Gamma_{ij}$.
Thus we have shown that, up to small corrections (of $\Delta\Gamma_{ij}$),
we in fact get out the same $\ham_\MF$ that we put in, so our
theory is self-consistent.
The only condition required for this to work was ~(\ref{eq:gamma_emp}).

Note $\Gamma^{(0)}$ and $\Gamma^{(2)}$ are 
of order $S \ln \varepsilon$, as will be 
explicitly verified analytically for the checkerboard lattice
(Sec.~\ref{sec:check_mft}) and the pyrochlore
(Sec.~\ref{heff_logdiv}).
The correction $|\Delta \Gamma_{ij}|$ in \eqr{eq:gamma_emp}
is an order of magnitude smaller than $\Gamma^{(2)}$ 
for all tractable values of $\varepsilon$.

If we had tried a different one-parameter form of 
variational Hamiltonian, where we add
$\pm\delta$ to the matrix elements $\Hvec_{ij}$ in a pattern other than
the one in
Eq.~(\ref{eq:variational}), the divergent $\Gamma_{ij}$ would indeed
be regularized, but the dominant
contribution would still be of the form~(\ref{eq:gamma_emp}), so 
self-consistency is lost: the output would not have the same as the input
The only one-parameter nearest-neighbor variational Hamiltonian which
is self-consistent is~(\ref{eq:variational}).

\subsubsection{Variational approach} \label{sec:var-approach}

The above recipe is perfectly valid, but our actual calculation
was done somewhat differently.
We diagonalized the $\ham_\var$ to find a variational wavefunction
$\Psi_\var(\varepsilon)$ and its
correlations $\{G_{ij}\}$, 
and computed an expectation $E_\MF(\varepsilon,S)$ [given by (\ref{eq:emf})].
We iteratively minimized $E_\MF(\varepsilon,S)$ 
with respect to $\varepsilon$ (for a given $S$), defining a unique
optimal value $\varepsilon=\varepsilon^*(S)$.
% for which the variational wavefunction $\Psi_\var(\varepsilon^*)$
% minimizes $E_\MF$.

It will be shown below (in Sec.~\ref{sec:check_mft} and ~\ref{heff_logdiv})
that $\varepsilon^*(S) \propto \ln{S}/S$.

\subsubsection{Equivalence of self-consistent and variational approaches}
\label{sec:selfcons-OK}

It remains to be justified that $\epsSC$, defined self-consistently,
should equal $\varepsilon^*$, defined by minimizing $E_\MF$. 
This is expected, since the decoupling is variationally based:
that is, a {\it full} variational optimization of $\ham_\MF$ 
with respect to all its parameters is equivalent to 
self-consistency with the decoupling form, by construction.
Thus, to the extent the full solution sticks within the subspace
defined by $\ham_\var$ (as we argued it did), the decoupling and
variational minimization (both within that subspace)
ought to agree with each other.

\begin{figure}
\begin{center}
\resizebox{\columnwidth}{!}{\includegraphics{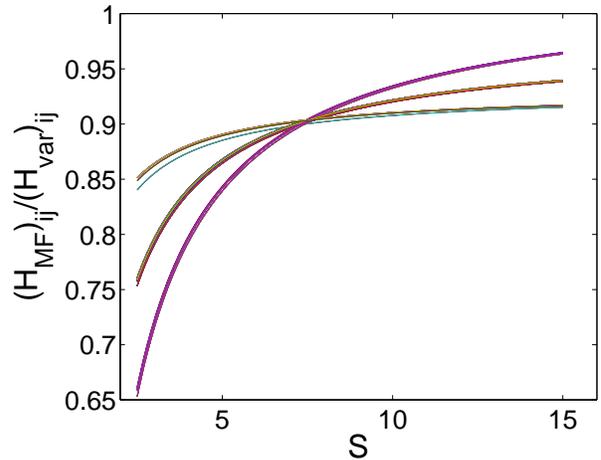}}
\end{center}
\caption{  \label{fig:self_const} \footnotesize (Color Online)
Self-consistency of the matrix elements.
We show the ratio of all nonzero elements of $\Hvec_\MF$ and
$\Hvec_\var(\varepsilon)$ for the state shown in Fig.~9(d) of
Ref.~\onlinecite{uh_harmonic}.
Here $\varepsilon$ is set to $0.1$.
Each line represents a particular $(ij)$ matrix element.
Up to symmetries of the configuration, there are $11$ unique matrix 
elements for this state, 
some of which are virtually indistinguishable in the plot.
All of the lines converge at $S^*(\varepsilon=0.1)=7.5$
(up to a deviation which is much smaller than $\varepsilon$).}
\end{figure}

The test for whether our result really is self-consistent is
that the diagonal elements~(\ref{eq:hmf_ii}) should be independent of $i$,
 and the off-diagonal elements~(\ref{eq:hmf_ij}) should depend solely on
$\eta_i \eta_j$.
Furthermore, we want $(H_\MF)_{ij}/(H_\var)_{ij}$  to be equal for all $i$, $j$
(for which $H_{ij}\ne 0$).
We indeed found (empirically) that this works when
$\varepsilon= \varepsilon^*(S)$, i.e.
[letting $S^*(\varepsilon)$ be the inverse relation to
$\varepsilon^*(S)$]
\begin{equation}\label{eq:variance_ratio}
\mathrm{variance}_{ij}
\left\{ \frac{(H_\MF(S^*(\varepsilon))_{ij}}{(H_\var(\varepsilon))_{ij}} \right\}
\ll \varepsilon\,.
\end{equation}
In Fig.~\ref{fig:self_const} we show an example of this for a particular state
and a particular value of $\varepsilon$.
The crossing defines $\epsSC$, in light of (\ref{eq:HMF-agrees}), 
but it is seen to happen exactly where $\varepsilon=\varepsilon^*$,
thus empirically confirming the equivalence.

\subsection{Scaling} \label{sec:scaling}

Within the harmonic theory of Ch.~\ref{sec:harmonic}, the fluctuations of
the spin deviation operators scale as
$\langle \sigma_i \sigma_j \rangle  = \OO(S)$
-- we omit  the  $x$ and $y$ component labels
in these schematic expressions -- 
and therefore we would na\"ively expect, from the spin-wave
expansion~(\ref{eq:ham_expand}), that 
\begin{equation} \label{eq:scaling_naive}
E_\harm = \OO(S)\,, \qquad
\langle \ham_\quart \rangle_{\rm naive} = \OO(1)\,.
\end{equation}
However, $\ham_\quart$ has an infinite expectation using the
unmodified ground-state wavefunction of $\ham_\harm$, 
since the fluctuations diverge.
Studies of the kagom\'e lattice~\cite{chubukov,chan,chan_thesis} have
taught us that, when anharmonic terms are treated self consistently,
spin fluctuations of \emph{divergent modes} are renormalized to
finite values.
In the kagom\'e case $\langle \sigma_i \sigma_j \rangle  = \OO(S^{4/3})$
and the scaling relations are
\begin{equation} \label{eq:scaling_kagome}
E_\harm = \OO(S)\,, \qquad
\langle \ham_\quart \rangle_{\rm kag} = \OO(S^{2/3})\,.
\end{equation}
Note that the harmonic energy is not rescaled because the frequency of divergent
zero modes is only $\OO(S^{2/3})$, which is negligible compared to non-zero
modes' $\OO(S)$ frequency.

One might expect the scaling (\ref{eq:scaling_kagome}) to carry through 
to the pyrochlore lattice as well\cite{kvale}.
%% However, it is important to observe that the dominant contribution to the
%% anharmonic term in $E_\MF$ comes from (the vicinity of) the divergent modes.
%, discussed previously in Sec.~\ref{sec:divergent}.
However, the divergent modes of the kagom\'e and the pyrochlore
are rather different:
in the kagom\'e, due to the anisotropy between in-plane and out-of-plane spin
fluctuations, \emph{all} zero modes are divergent modes, so the
kagom\'e divergent modes span the entire Brillouin zone.
In the pyrochlore, on the other hand, the divergent modes reside only along
lines in the Brillouin zone, hence the divergences (coming from these lines'
vicinity) are weaker.
%% However, it is important to observe that the dominant contribution to the
%% anharmonic term in $E_\MF$ comes from (the vicinity of) the divergent modes.
Below [see Eqs.~(\ref{eq:gij_r_check}) and (\ref{eq:gij_r})]
we shall find that this leads
to \emph{logarithmic} renormalization of the divergent fluctuations
$\Gamma_{ij}  =  \OO (S \ln{S})$, resulting in scaling
\begin{equation} \label{eq:scaling_pyrochlore}
\langle \ham_\quart \rangle \equiv E_\MF-E_\harm= \OO((\ln{S})^2)\,.
\end{equation}

The singularity of the divergent modes' fluctuations, away from 
$\qvec = \Ovec$, is cut off by the variational parameter $\varepsilon$.
At $\qvec = \Ovec$,
the divergence of $\langle \sigma_i \sigma_j \rangle $ would be
preserved, due to the physical Goldstone mode $\vvec_G$,
but the Goldstone mode's contribution to $\Gamma_{ij}$ vanishes such that the
Goldstone mode does not contribute to the energy at any order in $1/S$.

Because it is technically difficult to deal with the divergence of
$G_{ij}(\qvec = \Ovec)$ we shall, for now,
retain both variational parameters.
Thus we will have a handle on the fluctuations until we
eventually take the limit $\delta \!\to\! -\varepsilon/4$.
[We find that $G_{ij}(\qvec=\Ovec) \sim 1/\sqrt{\varepsilon+4 \delta}$, so
that $\varepsilon+4 \delta$ must be chosen to be positive.]

\section{Checkerboard lattice}\label{sec:checker}

As a warm-up to the pyrochlore lattice problem, we first consider the
same model on the closely related, two-dimensional checkerboard lattice.
This case is more tractable, in that some expressions have a simple form
which could not (or should not) be written out analytically in the
pyrochlore case.  

The checkerboard lattice (see Fig.~\ref{fig:checkerboard}) 
can be viewed as $\{001\}$ projection
of the pyrochlore lattice, and is often called the \emph{planar pyrochlore}.
The lattice structure is a square lattice with
primitive vectors $(1,1)$, $(1,-1)$
and two sublattices corresponding to
basis vectors $(-1/2,0)$ and $(1/2,0)$.
We refer to the crossed squares as ``tetrahedra'' in analogy with the
pyrochlore lattice, and we refer to any two sites within a tetrahedron 
as ``nearest neighbors'' regardless of the actual bond length.

Since the checkerboard lattice, as the pyrochlore,
is composed of corner sharing tetrahedra,
the derivation of Ch.~\ref{sec:setup} remains valid.
Note that we assume that all of the couplings within a tetrahedron are 
equal, even though in the checkerboard lattice,
the various bonds are not related by lattice symmetries. 
Since the shortest loop in the checkerboard lattice is a square, 
the effective harmonic Hamiltonian for this lattice has the same form as
the pyrochlore harmonic effective Hamiltonian~(\ref{eq:heff_harm}),
with the addition of a dominant term $K_4 \Phi_4$, 
with $K_4 < 0$.\cite{tcher_check,uh_harmonic}

Thus, the harmonic ground states of the checkerboard lattice consist of all
the zero-flux states, i.e., states with positive flux in all square plaquettes.
Similar to the pyrochlore case, this is a family of states that are exactly
degenerate to harmonic order, and in this case the residual entropy is
$\OO(L)$, where $L$ is the linear dimension of the system.\cite{uh_harmonic}
But since lattice does not respect the full symmetry of the tetrahedron,
the selection effect of the anharmonic terms turns out quite different
(and essentially trivial) as compared to the pyrochlore case.

\subsection{The checkerboard ($\pi$,$\pi$) state}\label{sec:check_pipi}

\begin{figure}
\begin{center}
\resizebox{\columnwidth}{!}{\includegraphics{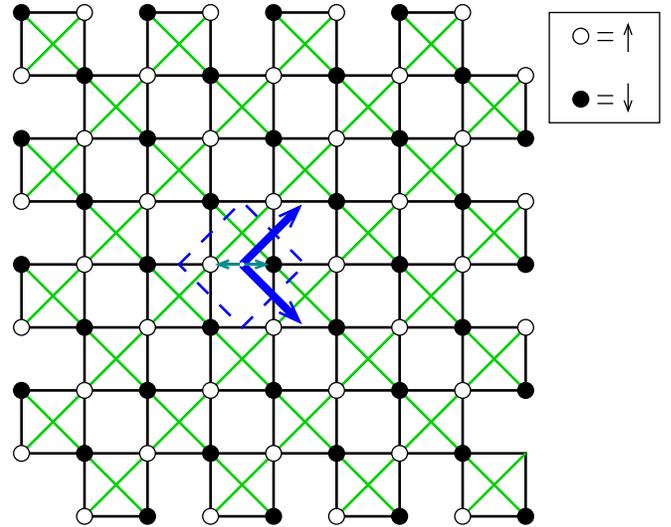}}
\end{center}
\caption{ \label{fig:checkerboard}\footnotesize (Color Online)
The checkerboard lattice $(\pi,\pi)$ state.
The primitive vectors are the diagonal arrows, and
the primitive unit cell is shown by the dashed square.
The small arrows represents the two basis vectors.
Here we show the $(\pi,\pi)$ state: open (closed) circles denote 
up (down) spins. Dark (light) colored lines denote AFM
(FM) bonds.}
\end{figure}

One of the checkerboard harmonic ground states is simple enough for the
diagonalization of the variational Hamiltonian~(\ref{eq:variational})
to be done analytically: the $(\pi,\pi)$ state depicted in
Fig.~\ref{fig:checkerboard}.
In this state, the diagonal bonds in each tetrahedron are
unsatisfied (FM), such that the symmetry of the lattice is conserved,
and the magnetic unit cell has two sites.

\subsubsection{Harmonic Hamiltonian for checkerboard} \label{sec:check_harm}

The Fourier transformed harmonic Hamiltonian
for the $(\pi,\pi)$ state is Eq.~(\ref{eq:harm_mat_q}), with
\begin{equation} \label{eq:ham_check}
\Hvec(\qvec)=
\left( \begin{array}{cc}
2 \cos^2  Q_{+} & 2 \cos Q_{+} \cos Q_{-}\\
2 \cos Q_{+} \cos  Q_{-} &2 \cos^2 Q_{-} \\
\end{array} \right) \,,
\end{equation}
where 
\begin{equation}
Q_\pm \equiv (q_x \pm q_y)/2\,.
\end{equation}
The spin-wave modes can be found
by diagonalizing the matrix $\etvec \Hvec(\qvec)$.\cite{uh_harmonic}
$\etvec$ is a diagonal matrix with elements $\{\eta_i\}$ along the diagonal
(in our case $\eta_1=1$, $\eta_2=2$).
Diagonalization of $\etvec \Hvec(\qvec)$ produces eigenmodes $\Vvec_\qvec$
and $\Uvec_\qvec$ for any wavevector $\qvec$
\begin{eqnarray}
\Vvec_\qvec^T &= \sqrt{\frac{2}{\alpha_\qvec}} (\cos Q_+, -\cos Q_-) \,,\qquad
\lambda_V &= \beta_\qvec \,, \nonumber \\
\Uvec_\qvec^T &=\sqrt{\frac{2}{\alpha_\qvec}} (\cos Q_-, -\cos Q_+) \,, \qquad
\lambda_U &= 0 \,,
\end{eqnarray}
satisfying the pseudo orthogonality condition
$\Vvec_\qvec^\dagger \etvec \Uvec_\qvec = 0$. 
The dispersions corresponding to $\Vvec_\qvec$ and $\Uvec_\qvec$, respectively
are
\begin{equation} \label{eq:check_disp}
\lambda_{\Vvec_\qvec} = \beta_\qvec \,, \qquad
\lambda_{\Uvec_\qvec} = 0 \,,
\end{equation}

Here we defined
\begin{eqnarray}
\alpha_\qvec &=& 2(\cos^2 Q_+ +\cos^2 Q_-)\,, \nonumber\\
\beta_\qvec  &=& 2(\cos^2 Q_+ - \cos^2 Q_-)\,.
\end{eqnarray}
%
%Note that Eq.~(\ref{eq:ham_check}) can be written as
%
%%\begin{equation}
%$\Hvec (\qvec)  =  
%\alpha_\qvec \etvec \Vvec_\qvec \Vvec_\qvec^T \etvec$. % \,.
%%\end{equation}
%
Thus, the ordinary spin-wave band has dispersion
$\hbar \omega_\qvec  =  2 S | \beta_\qvec |$,
and the zero point energy can be easily calculated 
\begin{equation}
E_\harm = \frac{1}{2} \sum_\qvec \hbar \omega_\qvec -N_sS =
N_s S\left(\frac{4}{\pi^2} -1\right)\,.
\end{equation}
The fluctuations of the spin deviation operators
($G_{lm}(\qvec) = \langle \sigma_\qvec^x(l) \sigma_{-\qvec}^x(m)\rangle$,
where $l$ and $m$ are sublattice indices) can be calculated 
from the spin-wave modes by Eq.~(\ref{eq:fluct})
\begin{equation}\label{eq:check_corrq}
\mathbf{G}(\qvec)=
\frac{S}{2 \beta_\qvec}
\left( \begin{array}{cc}
\alpha_\qvec & -\gamma_\qvec\\
-\gamma_\qvec& \alpha_\qvec 
\end{array} \right) \,,
\end{equation}
where $\gamma_\qvec \equiv 4\cos Q_+ \cos Q_-$, so that
$\alpha_\qvec = \sqrt{\beta_\qvec^2 + \gamma_\qvec^2}$.
Eq.~(\ref{eq:check_corrq}) shows that the
fluctuations diverge wherever $\beta_\qvec$ vanishes,
i.e., along the lines in the Brillouin zone $|Q_+|=|Q_-|$, which turn out to be
$q_x = 0$ or $q_y = 0$.

\subsubsection{Anharmonic energy} \label{sec:check_mft}

The variational Hamiltonian for the $(\pi,\pi)$ checkerboard state is of the
form~(\ref{eq:harm_mat_q}) with the matrix~(\ref{eq:variational}) given by
\begin{equation}
\Hvec_\var (\qvec) = \alpha_\qvec \etvec \Vvec_\qvec \Vvec_\qvec^T \etvec +
\delta \alpha_\qvec  \Vvec_\qvec \Vvec_\qvec^T + \varepsilon \openone \,,
\end{equation}
Diagonalizing $\etvec \Hvec_\var (\qvec)$, and keeping only the first order
terms in $\delta$, $\varepsilon$ results in
$\omega_\qvec$ of order $\sqrt{\varepsilon}$, $\sqrt{\delta}$ along
the divergence lines defined by $\beta_\qvec = 0$, and a linear
(in $\varepsilon$,$\delta$) correction to $\omega_\qvec$ away from these lines.

The fluctuations of the variational Hamiltonian are now:
\begin{equation}
%\langle \svec^{x/y}_\qvec (\svec^{x/y}_{-\qvec})^\dagger \rangle 
\mathbf{G}(\qvec) 
=\frac{S}{2D_\qvec(\varepsilon,\delta)}
%\nonumber 
%\\ &&\times
\left( \begin{array}{cc}
\alpha_\qvec (1+\delta) + 2 \varepsilon &
-\gamma_\qvec(1-\delta) \\
-\gamma_\qvec(1-\delta) &
\alpha_\qvec (1+\delta) + 2 \varepsilon
\end{array} \right) \,.
\end{equation}
Here we defined, for conciseness
\begin{equation}
D_\qvec(\varepsilon,\delta) \equiv \sqrt{\beta_\qvec^2(1-\delta)^2 +
4 (\alpha_\qvec +\varepsilon)(\alpha_\qvec \delta +\varepsilon)}\,.
\end{equation}
The fluctuations diverge (for nonzero $\varepsilon$) only if
$\beta_\qvec = 0$ \emph{and} $\alpha_\qvec \delta  +  \varepsilon = 0$.
If we take $\delta  \!\to\! -\varepsilon/4$, to conserve the symmetries of the 
original Hamiltonian, we find one divergent mode: the $\qvec = \Ovec$
Goldstone mode.

In order to calculate the mean field energy~(\ref{eq:emf}), 
we are interested in combinations of the diagonal (on-site) and off-diagonal
(nearest neighbor) fluctuations of the form $\Gamma_{ij}$.
We can write this as a sum over Fourier modes
\begin{equation}
\Gamma_{ij} =\frac{1}{N_M} \sum_\qvec \Gamma_{ij} (\qvec)\,,
\end{equation}
with $\Gamma_{ij}(\qvec)$ defined as
\begin{equation} \label{eq:corrdiffq}
\Gamma_{ij}(\qvec) \equiv G_{l_i l_i}(\qvec)-\eta_i \eta_j G_{l_i l_j}(\qvec)
\cos \xivec_{ij}\cdot \qvec\,.
\end{equation}
Here $l_i$, $l_j$ are the sublattice indices of $i$ and $j$, respectively,
$\xivec_{ij}$ is the vector connecting the two sites.
$N_M$ is the number of points in the Brillouin zone, i.e. the number of sites
in the magnetic lattice.

\begin{figure}
\begin{center}
(a) \resizebox{!}{6cm}{\includegraphics{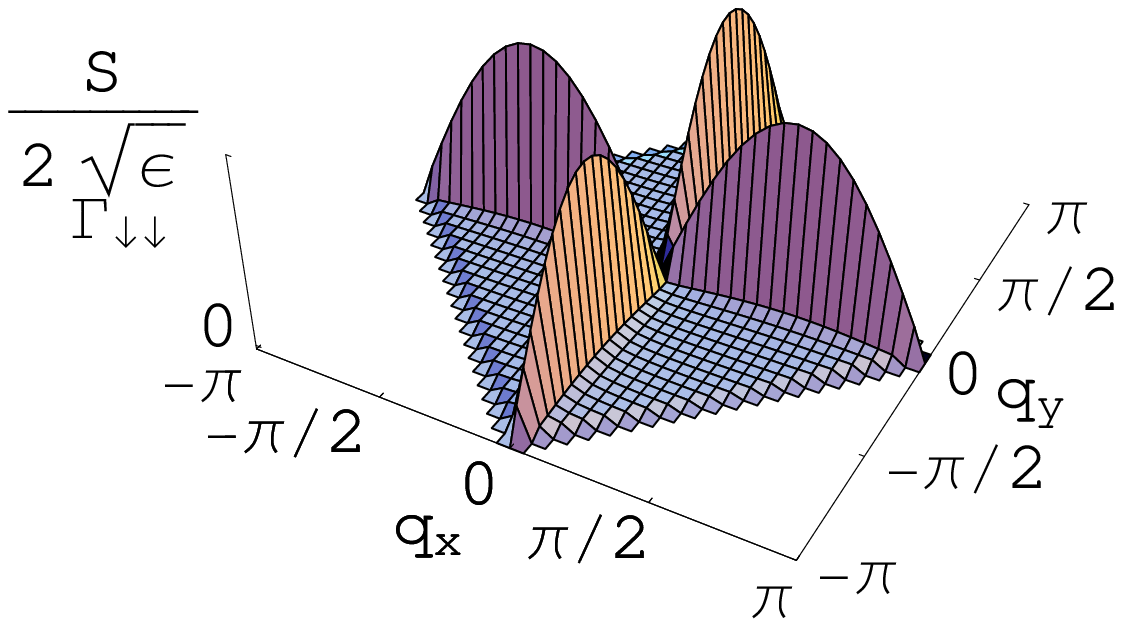}}\\
(b) \resizebox{!}{6cm}{\includegraphics{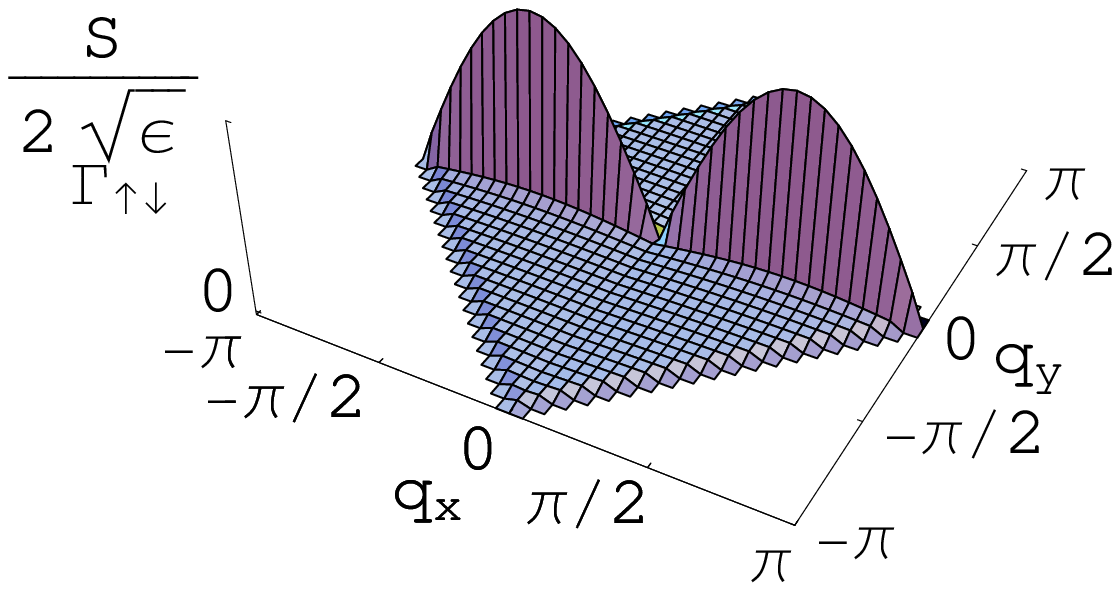}}
\end{center}
\caption{\label{fig:g_ij} \footnotesize  (Color Online)
Bond variables in the Brillouin zone of the $(\pi,\pi)$ checkerboard state.
(a) $\Gamma_{ij}$ for two neighboring sites on the same sublattice.
(b) $\Gamma_{ij}$ for two neighboring sites with $\eta_i \eta_j  = -1$. 
In the case shown, the $(ij)$ bond is along the $x$ axis.
The analytic forms of the functions are given in Eq.~(\ref{eq:corrdiff_uu})
and Eq.~(\ref{eq:corrdiff_ud}), respectively.}
%For both plots, the singularity of $\Gamma_{ij}$ goes as $S/2\sqrt{\varepsilon}$,
%and there is no divergence at $\qvec=\Ovec$.
%The width of the singularity goes as $1/\sqrt{2\varepsilon}$.
\end{figure}

In this case we obtain, for two neighboring sites on the same sublattice
\begin{eqnarray} \label{eq:corrdiff_uu}
\Gamma_{\uparrow \uparrow}(\qvec)&=&
 \frac{S}{D_\qvec(\varepsilon,\delta)}
 [\alpha_\qvec (1+\delta) + 2 \varepsilon ]\sin^2 Q_+  \,,\\
 \label{eq:corrdiff_dd}
\Gamma_{\downarrow \downarrow}(\qvec)&=&
 \frac{S}{D_\qvec(\varepsilon,\delta)}
 [\alpha_\qvec (1+\delta) + 2 \varepsilon ]\sin^2 Q_-  \,.
\end{eqnarray}
Here we used $\Gamma_{\uparrow \uparrow}(\qvec)$
[shown in Fig.~\ref{fig:g_ij}(a)] for $\Gamma_{ij}(\qvec)$, where both $i$ and $j$
are on the up-spin sublattice (and similarly for $\Gamma_{\downarrow\downarrow}$.
For neighboring sites on different sublattices,
we obtain [see Fig.~\ref{fig:g_ij}(b)]
\begin{equation} \label{eq:corrdiff_ud}
\Gamma_{\uparrow \downarrow}^{x/y}(\qvec)
= \frac{S}{2D_\qvec(\varepsilon,\delta)}
 [\alpha_\qvec (1+\delta) + 2 \varepsilon - \gamma_\qvec(1-\delta)\cos q_{x/y}]
 \,,
\end{equation}
where $\Gamma_{\uparrow \downarrow}^x$ ($\Gamma_{\uparrow \downarrow}^y$)
is the bond variable for a bond oriented along the $x$ ($y$) axis, connecting
an up-spin and a down-spin.
Note that Eqs.~(\ref{eq:corrdiff_uu}),(\ref{eq:corrdiff_dd}),
and~(\ref{eq:corrdiff_ud}) do not diverge at any value of $\qvec$
for $\varepsilon + 4\delta = 0$.
Thus, we have regularized the fluctuations, and retained only one variational
parameter.
%(ii) $\Gamma_{ij}(\qvec) \ge   0$, as must always be the case.
Since all sites are related by symmetry in this state,
$\Gamma_{ij} = \Gamma_{ji}$.
Furthermore $\Gamma_{\uparrow \uparrow}(\qvec)$ and $\Gamma_{\downarrow\downarrow}(\qvec)$
are related by a rotation of the Brillouin zone, and the real space correlations
will be the same upon integration over the Brillouin zone.

As we can see in Fig.~\ref{fig:g_ij}, the divergent lines for
$\Gamma_{\uparrow\uparrow}(\qvec)$ and $\Gamma_{\downarrow\downarrow}(\qvec)$
are both major axes, whereas $\Gamma_{\uparrow\downarrow}^x(\qvec)$ and
$\Gamma_{\uparrow\downarrow}^y(\qvec)$
only diverge along the $y$ and $x$ axes, respectively.
Along the divergent lines, where $\beta_\qvec = 0$ and
$\alpha_\qvec = |\gamma_\qvec| = 4 \cos^2 Q_+$, 
the values of the bond variables are, asymptotically
$\Gamma_{ij}(\qvec) = S|\sin 2 Q_+| /2\sqrt{\varepsilon}$.
Away from the divergence line,
\begin{equation} \label{eq:gij_perp_check}
\Gamma_{ij}(\qvec)\approx \frac{S|\sin 2Q_+|}{2\sqrt{\varepsilon + 4 q_\perp^2}}
\,, 
\end{equation}
where $q_\perp \ll 1$ is transverse to the divergence line.
Upon integration of~(\ref{eq:corrdiff_uu}),~(\ref{eq:corrdiff_dd}),
and~(\ref{eq:corrdiff_ud}) over the Brillouin zone,
the result is a logarithmic singularity in the fluctuations: 
\begin{equation} \label{eq:gij_r_check}
\Gamma_{\uparrow\uparrow}=\Gamma_{\downarrow\downarrow}=
-\frac{4S}{\pi^2} \ln \varepsilon +\OO(\varepsilon) =
2 \Gamma_{\uparrow\downarrow} + \OO(\varepsilon) \,.
\end{equation}
Observe that, in the notation of (\ref{eq:gamma_emp}),
$\Gamma_{\uparrow\uparrow}=
\Gamma_{\downarrow\downarrow}=
\Gamma^{(0)}+\Gamma^{(2)}$ and
$\Gamma_{\uparrow\downarrow}=
\Gamma^{(0)}-\Gamma^{(2)}$, so the ratio 2
in Eq.~(\ref{eq:gij_r_check}) is equivalent to
the ratio 3 in (\ref{eq:gamma1to3}).
These fluctuations $\{ \Gamma _{ij} \}$, divergent  
as $\ln \varepsilon$, enter 
quadratically into the anharmonic term of
Eq.~(\ref{eq:emf}) for the mean field energy $E_\MF$
(The divergent part of the harmonic contribution, linear in
$\{ \Gamma_{ij} \}$, cancels as was noted in Sec.~\ref{sec:Gamma-form}.)

\begin{eqnarray}
E_\MF&&=E_\harm + S\times\OO(\varepsilon) -
\sum_{\langle ij \rangle} \eta_i \eta_j (\ln{\varepsilon})^2+
\OO(\varepsilon \ln{\varepsilon})
\nonumber \\
=&&
E_\harm +S\times\OO(\varepsilon)+\frac{4(\ln{\varepsilon})^2}{\pi^4}+
\OO(\varepsilon \ln{\varepsilon})\,.
\label{eq:check_emf_e}
\end{eqnarray}

Minimizing~(\ref{eq:check_emf_e}) with respect to $\varepsilon$,
for a given $S\!\gg\! 1$ (ignoring the subdominant last term),
we obtain $\varepsilon^*(S) \propto \ln{S}/S$
and therefore the quartic energy $E_\quart  \equiv  E_{MF}  -  E_\harm$
is quadratic in $\ln{S}$.
We remark that due to the logarithmic singularity, in a numerical calculation 
one would expect it to be hard to distinguish between terms
of order $\OO((\ln{\varepsilon})^2)$, $\OO(\ln{\varepsilon})$, and $\OO(1)$
for numerically accessible values of $\varepsilon$. 
Nevertheless, since we are doing a large-$S$ expansion, we are
mostly interested in the asymptotic behavior.

\SAVE{We could do the same thing for the $(\pi,0)$ state and compare.but:
(i) expressions are not as neat.
(ii) Magnetic unit cell is larger with two non-zero bands. Or
(iii) Same magnetic unit cell, using bond order, but then gauge transformations
have an effect of moving modes from one $\qvec$ to another,
so it's hard to compare.
}

\subsection{Anharmonic ground state selection} \label{sec:checker_select}

Now that we looked at the checkerboard $(\pi,\pi)$ state, what can be said 
about the anharmonic selection in the checkerboard lattice?
The harmonic ground states in the checkerboard are the zero-flux state:
all of the states that have a positive product over $\eta_i$ around all
square plaquettes.

In this section, we shall first find the ordinary spin-wave modes
(ignoring the generic zero modes, which are the same for all states), and
then focus on the divergent modes to predict which state is favored.
Next, we show some numerical evidence to support are prediction.

\subsubsection{Spin-wave modes for a generic harmonic ground state}

In order to understand the leading order term in the anharmonic energy, 
we restrict our discussion to the correlations due to divergent modes.
We would like to derive an expression for $\Gamma_{ij}$,
for any zero-flux state.

We start by explicitly finding the ordinary spin-wave modes of the harmonic
Hamiltonian~(\ref{eq:harm_mat}).
Recall that the divergent modes are a subset (of measure zero) of the ordinary
modes. Since we expect the divergent and nearly-divergent modes to dominate
the fluctuations, we shall later limit ourselves to 
ordinary modes in the vicinity (in $\qvec$-space) of the divergent modes.

As we saw in Sec.~\ref{sec:ordinary}, any ordinary mode
$\vvec_m$ can be written [Eq.~(\ref{eq:ord_modes})] in terms of a vector
$\uvec_m$, of length $N_s/2$, living on the centers
of ``tetrahedra''. In the checkerboard case, these correspond to square lattice 
sites. $\{\uvec_m\}$ satisfy the spin-wave equation~(\ref{eq:sw_diamond}),
which can easily be solved by an ansatz 
\begin{equation}\label{eq:check_umode}
u_\qvec (\alpha) = \nu_\alpha \sqrt{\frac{2}{N_s}}
e^{i \qvec\cdot \rvec_\alpha}\,,
\end{equation}
with $\nu_\alpha= \pm 1$ (to be determined).
Plugging this into~(\ref{eq:sw_diamond}),
we obtain, for any $\alpha$
\begin{equation} \label{eq:check_ord_lambda}
\lambda_\qvec = \half
{\sum^\ord}_{\beta {\rm~n.n.~of~}i}
\eta_{i(\alpha\beta)} \nu_\alpha \nu_\beta
e^{i\qvec\cdot(\rvec_\beta - \rvec_\alpha)} \,.
\end{equation}
As always, ``$\ord$'' denotes a quantity limited to contributions
from ordinary modes.
In order for the right-hand-side of~(\ref{eq:check_ord_lambda}) to be
independent of $\alpha$, we choose
\begin{equation} \label{eq:nu_alpha}
\nu_\alpha \nu_\beta = \eta_{i(\alpha\beta)}\,.
\end{equation}
It is easy to check that for (only) zero-flux states, the signs of $\{\nu_\alpha\}$
can be chosen consistently so that~(\ref{eq:nu_alpha}) is satisfied.
(Note there is no need to assume the state is periodic.)
% Note that this solution is valid for any arbitrary zero-flux state, with no
% assumption of periodicity.
Thus we obtain, from~(\ref{eq:check_ord_lambda}), that for any checkerboard 
lattice zero-flux state, the dispersion is
\begin{equation}
\lambda_\qvec = 2\cos q_x \cos q_y \,.
\end{equation}
Note that here $q_x$ and $q_y$ are shifted by $(\pi/2,\pi/2)$ compared to 
Eq.~(\ref{eq:check_disp}) [for the $(\pi,\pi)$ state].
This dispersion is shared by all of the harmonic ground states of the 
checkerboard.

The (normalized) checkerboard-lattice ordinary spin-wave modes are,
using ~(\ref{eq:check_umode}) in~(\ref{eq:ord_modes}),
thus
\begin{equation}\label{eq:check_ordmodes}
v_\qvec(i) = \eta_i \frac{1}{\sqrt{N_s}}
\sum _{\alpha: i\in\alpha}
\nu_\alpha e^{i\qvec\cdot \rvec_\alpha} \, .
% \left[\nu_{\beta(i)} e^{i\qvec\cdot \rvec_{\alpha(i)}}+
% \nu_{\alpha(i)} e^{i \qvec\cdot\rvec_{\beta(i)}} \right]\,,
\end{equation}

%% The ``norm'' $\vvec_\qvec^\dagger \etvec \vvec_\qvec$
%% that goes into the denominator of 
%% Eq.~(\ref{eq:fluct}), is equal to 
%
%% \begin{equation}
%% \vvec_\qvec^\dagger \etvec \vvec_\qvec = \frac{1}{N_s}
%% \sum_{\langle \alpha \beta \rangle} [\eta_{\alpha\beta} 
%% + \cos\qvec\cdot(\rvec_\alpha - \rvec_\beta)] = \lambda_\qvec \,.
%% \end{equation}
%

The first term above vanishes upon summing over the lattice.

\subsubsection{Divergent correlations}

From (\ref{eq:check_ordmodes}), 
we can calculate the correlations due to ordinary modes,
using~(\ref{eq:fluct}) and~(\ref{eq:eta-norm-ord})
\begin{equation}\label{eq:check_Gord}
{G^\ord}_{ij} =  \eta_i \eta_j 
\sum_{\alpha: i\in\alpha} \sum_{\beta: j\in\beta} \nu_\alpha \nu_\beta
\tilde{g}_{\alpha\beta}
\end{equation}
where
\begin{equation}\label{eq:check_diamondcorr}
\tilde{g}_{\alpha\beta} \equiv
\frac{S}{2N_s}  \sum_\qvec 
{\frac{\cos \qvec\cdot(\rvec_\alpha-\rvec_\beta)} {|\lambda_\qvec|}} 
%% 2 \eta_i \cos{\qvec\cdot(\rvec_{\alpha(i)} - \rvec_{\beta(i)}}) \right]
\,.
\end{equation}
is manifestly independent of which (zero-flux) state we have.
Remember sum (\ref{eq:check_Gord}) has four terms;
in the limit of a large system,
the sum (\ref{eq:check_diamondcorr}) converts to an integral
in the standard fashion.
This is a special case of 
Appendix \ref{app:ord-calc}:
\eqr{eq:check_Gord} corresponds to \eqr{eq:Gord-uu},
and \eqr{eq:check_diamondcorr} corresponds to \eqr{eq:g_ab}
with $\tilde{g}_{\alpha \beta}= \nu_\alpha \nu_\beta g_{\alpha \beta}$.

At this point it appears that we have a problem. The integrand 
in (\ref{eq:check_diamondcorr}) diverges 
for any $\qvec$ along the divergence lines, and therefore we, of course, 
the correlations $G_{ii}$, $G_{ij}$ diverge for the 
unperturbed harmonic theory.
However, we have found that an adequate regularization scheme, such 
as the variational Hamiltonian~(\ref{eq:variational}), cuts off the singularity
and results in a logarithmic dependence. 
In particular, we have seen that, for the $(\pi,\pi)$ state,
$\frac{1}{N_s} \sum_\qvec (1/|\lambda_\qvec|)$ can be
replaced by a constant $C(\varepsilon)$ which is logarithmic in $\varepsilon$.
Since the dispersion of $\lambda_\qvec$ is the same for any 
zero-flux state, then $C(\varepsilon)$
can be assumed to be the same for all of the harmonic ground states.

\SAVE{CLH [9/08] believes the arguments about integrals (below)
could be rephrased in terms of $g_1$, $g_2$, and $g_{\sqrt 2}$. 
This would usefully disentangle factors that just depend on the structure
of the integrand and $\lambda_\qvec$, from those that depend
on the particular value of $\eta_i\eta_j$.}

Without loss of generality, suppose site 
$i$ is on the bond between diamond sites $\alpha$ and $\beta$ and
$j$ is shared by $\alpha$ and $\beta'$.
Plugging this into Eq.~(\ref{eq:corrdiff}) and using the
relation~(\ref{eq:nu_alpha}), we find the bond variables
\begin{eqnarray} 
{\Gamma^\ord}_{ij} &=& {G^\ord}_{ii}-\eta_i \eta_j {G^\ord}_{ij} \\ &=&
\frac{1}{N_s} \sum_\qvec \frac{S}{2|\lambda_\qvec|}
\big[ 1
-\eta_i \eta_j \cos{\qvec\cdot(\rvec_{\beta}-\rvec_{\beta'})}
\nonumber \\ &&\nonumber
+ \eta_i  \cos{\qvec\cdot(\rvec_{\beta} - \rvec_\alpha)} 
- \eta_j  \cos{\qvec\cdot(\rvec_{\beta'} - \rvec_\alpha)} \big]
\,.
\end{eqnarray}
The last two terms in this expression are identically $0$ 
(since the sum is odd in $\qvec$), and thus
\begin{equation}
{\Gamma^\ord}_{ij} = 
\frac{1}{N_s} \sum_\qvec \frac{S}{2|\lambda_\qvec|}
\big[ 1
-\eta_i \eta_j \cos{\qvec\cdot(\rvec_{\beta}-\rvec_{\beta'})}\big]
\,.
\end{equation}
Assuming that the anharmonic selection is solely due to nearly
divergent modes, we would like to focus on the vicinity  of the 
divergence lines in the 
Brillouin zone: $q_x \approx \pm \pi/2$ and $q_y \approx \pm \pi/2$.

If the bond $(ij)$ is diagonal,
$\rvec_{\beta} \!-\! \rvec_{\beta'} \!=\! (\pm 2,\pm2)$,
and the integral of the second
term over any of the divergence lines is identically
zero.\footnote{To see this, set $q_x=\pi/2+\Delta q_x$, and the second term is
$$
\eta_i \eta_j\left[
\left(\frac{1}{|\Delta q_x|}-2|\Delta q_x|\right)\mathrm{sgn}(\cos q_y)
- \frac{\sin q_y }{|\cos q_y|} \right]\,,
$$
which, upon integration over $q_y$ vanishes for arbitrarily small $\Delta q_x$.}
On the other hand, for a bond in the  $\mathbf{\hat{x}}$ ($\mathbf{\hat{y}}$)
direction, the bond term in the bracket is $+\eta_i \eta_j$ for
$\qvec=(\pm \pi/2,q_y)$ ($\qvec=(q_x,\pm \pi/2)$) and $0$ otherwise.

Thus we find 
\begin{equation} \label{eq:check_gij_div_case}
\Gamma_{ij} \approx \left\{ \begin{array}{ll}
SC(\varepsilon)  & (ij) \mbox{ diagonal bond}
\,, \\
SC(\varepsilon) (1+\half \eta_i \eta_j) & (ij)
\mbox{ $\mathbf{\hat{x}}$ or $\mathbf{\hat{y}}$ bond}\,, \\
0 & \mbox{otherwise} \,,
\end{array} \right.
\end{equation}

Comparing to~(\ref{eq:gamma_emp}), we
see that $\Gamma^{(0)} = S C(\varepsilon)$ while
$\Gamma^{(2)} = {\frac 1 2} \Gamma^{(0)}$ 
 on $\mathbf{\hat{x}}$ or $\mathbf{\hat{y}}$ bonds,
but zero on diagonal bonds;
the form is modified from ~(\ref{eq:gamma_emp}) owing to
the anisotropy of the ``tetrahedron'' in the checkerboard lattice
(i.e., the inequivalence of the two kinds of bond.)

Eq.~(\ref{eq:check_gij_div_case}) is by no means an exact result. 
We have made the following approximations in obtaining it:
(i) Neglecting modes away from the divergence lines.
This assumption is innocuous for large $S$, since the correlations
are dominated by the vicinity of divergent modes. \\
(ii) Neglecting all generic zero modes. In the checkerboard lattice, 
these modes, close to the divergence lines, can be shown to closely mimic 
the behavior of the ordinary modes, and will essentially increase
$C(\varepsilon)$ by a factor of $2$ (see Appendix~\ref{sec:app-genericzero}).

(iii) Ignoring any additional effects due to  the regularization scheme.
Although this assumption is not a priori justified, we would like,
as a first order approximation, to work with the 
bare harmonic Hamiltonian rather than the variational one, since it is
easier to deal with analytically.
We do not expect the regularization to qualitatively change the results we
discussed in the following.

\subsubsection{Single tetrahedron}\label{sec:check_single}

\begin{figure}
\begin{center}
\resizebox{!}{2.5cm}{\includegraphics{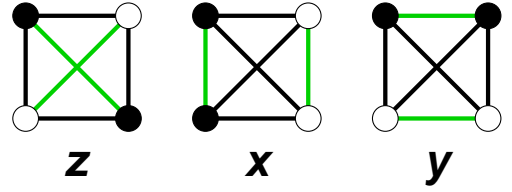}}
\end{center}
\caption{\label{fig:check_polarization} \footnotesize (Color Online)
The three possible polarization axes for a single tetrahedron.}
\end{figure}

To find the leading order quartic energy for a generic state, we consider
the three possible bond configurations for a single tetrahedron, which
can be viewed as three
\emph{polarization axes}:\cite{isakov,clh_polarization} $z$
(where all tetrahedra are oriented as in the $(\pi,\pi)$ state), $x$ and $y$ 
(see Fig.~\ref{fig:check_polarization}).\footnote{In
Ref.~\onlinecite{tcher_check}
the polarization axis of checkerboard
tetrahedra was denoted by a color Potts variable.}

Summing up the contributions, we obtain, for a single $z$ polarized tetrahedron:
\begin{equation}
E_\quart^\boxtimes = \frac{1}{S^2} \sum_{\langle ij\rangle \in \boxtimes}
\eta_i \eta_j \Gamma_{ij}^2 \approx
C(\varepsilon)^2 \,.
\end{equation}
On the other hand, for $x$ or $y$ polarization we find
\begin{equation}
E_\quart^\boxtimes \approx
2 C(\varepsilon)^2 \,.
\end{equation}
Note that in all cases $\sum \eta_i \eta_j \Gamma_{ij}^{(m)}  \approx  0$
to leading
order, since the divergent modes do not contribute to the harmonic part of
$E_\MF$ in~(\ref{eq:emf}).

Thus we found that the divergent contribution to the quartic energy is 
twice as large for $x$ or $y$ polarization as it is for $z$ polarization.
It follows that the effective Hamiltonian has the simplified form
\begin{equation}\label{eq:heff_check}
E_\quart^\eff = N_s[A(S) - B(S) \rho_z]\,,
\end{equation}
with $B(S)\approx A(S)/2$.
Therefore the $(\pi,\pi)$ state, in which all tetrahedra are $z$ polarized,
would be favored over all other zero-flux states, and thus
is the \emph{unique} ground state for the checkerboard lattice.

\subsubsection{Numerics for full lattice}

\begin{figure}
\begin{center}
\resizebox{!}{7cm}{\includegraphics{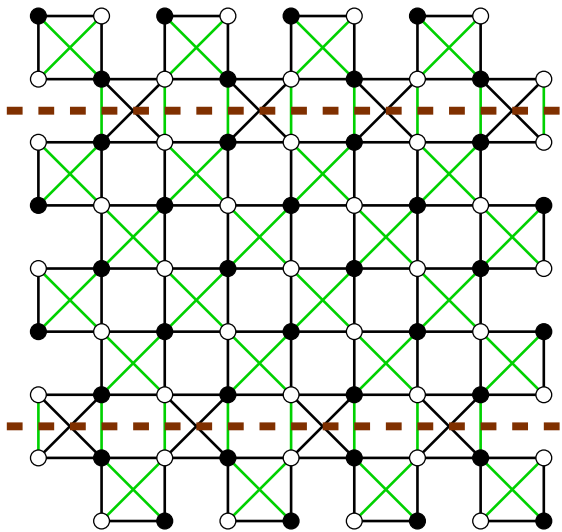}}
\end{center}
\caption{\label{fig:checkerboard_gauge}
\footnotesize (Color Online)
A checkerboard lattice harmonic ground state.
This state was constructed by flipping the bonds that cross each of the
two dashed horizontal lines.  }
\end{figure}

To confirm Eq.~(\ref{eq:heff_check})
on the anharmonic selection among harmonic checkerboard
ground states, we constructed various such states on a $8  \times  8$ cell
(see Fig.~\ref{fig:checkerboard_gauge}) in the following way:
we started from the $(\pi,\pi)$ state. 
There are $8$ horizontal lines, that each go through the centers of $4$ 
tetrahedra (dashed lines in Fig.~\ref{fig:checkerboard_gauge}).
We choose any of the $2^8$ subsets of these $8$ lines, and
change the sign of $\eta_i \eta_j$ on \emph{every} (vertical or diagonal)
bond that crosses one of the chosen horizontal lines.
It is easy to check, that each of these $2^8$ transformations
is a valid gaugelike transformation, since it does not violate the tetrahedron 
rule nor does it change the flux through any square plaquette.
It turns out that of the $2^8$ that can be obtained, only 
$32$ are unique by lattice symmetry.
Note that the construction of states, as well as our calculation, is based
on bond-order,\cite{uh_thesis} and thus we need not worry about
flipping an odd number of lines of this structure.~\footnote{
%%%%%%%%%%%%%%
It should be noted, however, that when we impose a gaugelike transformation
e.g. along a horizontal line, we are forced to change the vertical 
boundary condition from periodic to antiperiodic or vice versa.}
%%%%%%%%%%%%%%
See Ref.~\onlinecite{uh_harmonic} for a detailed discussion of gaugelike
transformations; for our purpose, it suffices to realize that each state
that we generate is a valid classical ground state with zero flux through 
each plaquette.

Whenever we flip a row of bonds, we change the polarization of four tetrahedra 
from the $z$ direction to the $x$ direction. 
Based on the arguments of the previous section, we expect that the 
leading order term in the quartic energy would be proportional to the number
of flipped rows.

\begin{figure}
\begin{center}
\resizebox{\columnwidth}{!}{\includegraphics{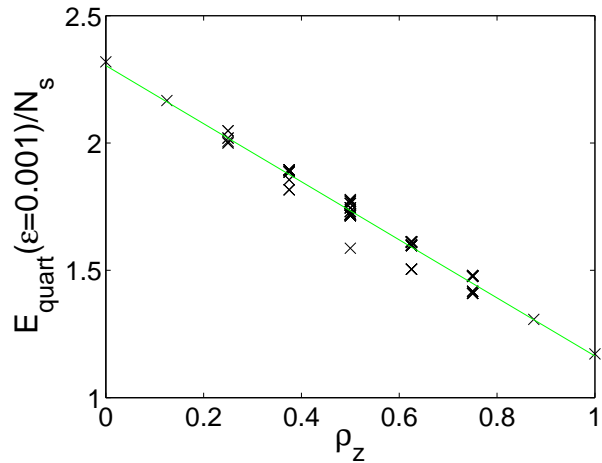}}
\end{center}
\caption{\label{fig:check_equart} \footnotesize   (Color Online)
Quartic energy for checkerboard lattice harmonic ground states.
The energy $E_\quart$ is shown for $\varepsilon = 0.001$, as
a function of the fraction of $z$ polarized tetrahedra,
for various checkerboard lattice harmonic ground states. }
\end{figure}

For each of these states, we calculate the quartic energy for a given value of 
$\varepsilon = 0.001$, integrating over $41 \times41$ points in the Brillouin
zone, equivalent to a system size of $328 \times 328$, which
is more than required to obtain good accuracy (see Sec.~\ref{sec:num} for more 
details about the numerical considerations).
The results are presented in Fig.~\ref{fig:check_equart}, as a function of
the fraction of $z$-polarized tetrahedra $\rho_z$. As expected we find:
(i) the quartic energy is, for the most part, linear in $\rho_z$.
(ii) the energy span is of order $4(\ln{\varepsilon})^2 / \pi^4$.
(iii) the ground state is the uniformly $z$ polarized $(\pi,\pi)$ state.
(iv) the quartic energy of the $(\pi,\pi)$ state is approximately half of
the energy of the uniformly $x$ polarized state.

Given the clear differences  in $E_\quart(\varepsilon,S)$ between the 
various harmonic ground states, we expect that the same ordering would
be conserved in the saddle point value $E_\quart(S)$
upon minimization with respect to $\varepsilon$.
Thus we can claim that the $(\pi,\pi)$ state is the zero-temperature, large-$S$,
ground state of the checkerboard lattice model.
This ground state is the same one found in large-$N$ calculations for
the large-$S$ limit.~\cite{bernier,uh_LN}
The effective quartic Hamiltonian has the form~(\ref{eq:heff_check})
with the coefficients $B(S) \propto  (\ln{S})^2$ and $A(S) \approx 2 B(S)$
to leading order in $S$.
We note that this effective Hamiltonian can be written in a more conventional
form, in terms of Ising products
\begin{equation}
E_\quart^\eff =
N_s A(S) - B(S) {\sum_{\langle ij\rangle }}^\times  \eta_i \eta_j\,,
\end{equation}
where $\sum^\times$ is a sum is over diagonal bonds only.

The result is not very surprising: although we set the Heisenberg couplings
to be the same for all bonds in the checkerboard lattice, there is no physical
symmetry between the diagonal bonds and the non-diagonal bonds
and therefore we should have expected to generate anharmonic terms consistent 
with the actual lattice symmetry.
Thus, unfortunately, this does not provide a guide to lattices where all bonds
in a tetrahedron are related by symmetry.

\section{Effective Hamiltonian for the pyrochlore}
\label{sec:num}

\SAVE{The Husimi cactus was first mentioned for frustrated systems
in  ``SPIN LIQUIDS ON THE HUSIMI CACTUS''
by P. Chandra and B. Dou\c{c}ot, J. Phys. A 27,  1541-1556   (1994). 
But I think it's enought to cite Ref.~\onlinecite{doucot}, 
``A semiclassical analysis of order from disorder'' (1998).}
%%%%%%%%%%%%%%%%%%%%%%%%%%%%%%%%%%%%%

We now turn our attention back to the pyrochlore lattice, where, due to the
large sizes of the magnetic unit cells of ground state candidates, it would
be challenging, at the least, to do analytic calculations 
(as were done for the checkerboard in Sec.~\ref{sec:checker}).
Since Sec.~\ref{sec:checker} explicitly worked out the details, for that case,
of implementing the self-consistent framework of Sec.~\ref{sec:mft}, we shall
not belabor steps which are roughly parallel.  However, the selection effects
themselves --- our ultimate motive --- are quite different now, since the
degeneracy is broken by {\it ordiinary} modes in the checkerboard case.

Our aim here is to calculate the quartic energy for a set of periodic
states, and gather the energies we have calculated to construct an effective 
Hamiltonian.
As seen in the harmonic theory of Ref.~\onlinecite{uh_harmonic}, and 
in the large-$N$ theory of Ref.~\onlinecite{uh_LN}, 
as well as the anisotropic perturbation theory of 
Refs.~\onlinecite{bergman} and \onlinecite{bergman-long}, 
it is natural that any non-trivial energy differences among states should be
represented as a sum over {\it loop} operators.
The effective Hamiltonian cannot take a local form:
the \emph{local} environments that all spins see are the same.
(Indeed, if we replaced the diamond lattice by a (loop-free) 4-coordinated 
lattice Bethe, so that our spin sites formed a ``Husimi cactus''~\cite{doucot},
then {\it all} Ising ground states would be equivalent by 
symmetry~\cite{uh_harmonic}.)

The numerical calculation is done as follows: for a given 
collinear classical ground state and a given value of 
$\varepsilon$ we diagonalize the Fourier
transform of the variational Hamiltonian~(\ref{eq:variational}),
keeping $\varepsilon + 4\delta$ infinitesimal.
We find the bond variable $\Gamma_{ij} (\qvec)$ for each wavevector
on a grid of Brillouin zone points, and sum over these points to obtain
$\Gamma_{ij}$ in real space.
Once we have calculated $E_\MF$ for many values of $\varepsilon$
(for a given collinear state), 
we can minimize it, for a given $S$, and find $E_\quart(S)$.
Our plan of action is to perform this numerical calculation of $E_\quart$
for a large database of collinear classical ground states and construct an
effective Hamiltonian.

\subsection{Logarithmic divergences} \label{heff_logdiv}

In performing the calculation, we find a distinct resemblance
to our findings on the checkerboard lattice:
There are divergent modes along the $x$, $y$, and $z$ axes in the Brillouin
zone,\cite{uh_harmonic} and these modes dominate the mean field quartic energy
(and have no contribution to the harmonic order energy).
The singularity of $\Gamma_{l_i l_j}(\qvec)$ is cut off,
along the divergence lines,
by a term of the order $S/\sqrt{\varepsilon}$.
The divergence peaks drop off to half of their maximum value at a ($\qvec$)
distance of order $\sqrt{\varepsilon}$, away from the divergence line.
This means that the grid of wavevectors that we use must be denser
in order to capture the effect of the divergent modes,
as $\varepsilon$ becomes smaller.
Thus, we need to sum of the order of $\varepsilon^{-3/2}$ points, 
to obtain good accuracy.
This limits the values of $S$ that we can do the calculation for,
and we have found no useful numerical tricks to get around it.
Nevertheless, we can get results over about two orders of magnitude of $S$, 
which can be extrapolated to the $S \!\to\! \infty$ limit.

Upon numerical integration, we find, that as in the
two-dimensional checkerboard lattice, the divergence of the fluctuations is
logarithmic
\begin{equation} \label{eq:gij_r}
\Gamma_{ij} \propto |\ln{\varepsilon}| + \OO(\varepsilon) \,.
\end{equation}
This numerical finding is somewhat surprising.
We would na\"ively expect that the bond variable $\Gamma_{ij}(\qvec)$ would
drop, away from the divergent lines, with a functional
form~(\ref{eq:gij_perp_check}), as in the checkerboard.
If so, as the transverse integration over $\qvec_\perp$ is now 
two-dimensional, the result would be a non-singular $\Gamma_{ij}$.

It turns out that this expectation is incorrect because the dispersion in the
direction perpendicular to the divergence line is strongly anisotropic.
For each value of $\qvec$ along the divergence line, there are two
particular independent eigendirections of $\qvec_\perp$.
For example, for a $\qvec = q_z \mathbf{\hat{z}}$ divergence, 
%corresponding to a linear combination of real-space divergent modes similar to
%the one shown in Fig.~\ref{fig:divergent}(b),
the eigendirections of $\qvec_\perp$ are $(1,1,0)$ and $(1,-1,0)$.
If we call unit vectors along these eigendirections
$\mathbf{\hat{e}_1}$ and $\mathbf{\hat{e}_2}$, then we find that
$\Gamma_{ij}\propto
1/\sqrt{\varepsilon+(q_\perp \cdot \mathbf{\hat{e}_1})^2}+
1/\sqrt{\varepsilon+(q_\perp \cdot \mathbf{\hat{e}_2})^2}$.
Integration over $\qvec_\perp$ results in the logarithmic dependence on 
$\varepsilon$ of~(\ref{eq:gij_r}), as in the checkerboard case.
In turn, as in Subsec.~\ref{sec:check_mft},
the logarithmic scaling of fluctuations in (\ref{eq:gij_r})
implies via (\ref{eq:vareps-selfcons}) that
\begin{equation} \label{eq:vareps-S}
 \epsSC(S) \propto \frac{\ln S}{S} \,.
\end{equation} 

Finally, we know the decoupled quartic energy in Eq.~(\ref{eq:emf}) 
is a sum over products $\Gamma_{ij} \Gamma_{ji}$, 
with $\Gamma_{ij}$ linear in $\ln{S}$; since the divergent parts linear in 
$\Gamma_{ij}$ cancel out [as noted before \eqr{eq:ham_var_dominant}],
the result is the anharmonic energy scales as $(\ln S)^2$, as
announced in \eqr{eq:scaling_pyrochlore}.

\subsection{Gauge invariant terms} \label{heff_gi}

For our database we calculated $E_\quart$ on a sample of classical ground
states (not all of them $\pi$-flux states), that we constructed by hand,
with unit cells ranging from 4 to 32 sites.
Two of these families consist of the zero-flux and $\pi$-flux states,
which have uniform $+1$ and $-1$ products around all hexagons,
respectively.
In the other three gauge families,
the hexagon fluxes are arranged in planes such
that within each plane the flux is uniform; we call these
the ``$000\pi$'', ``$0\pi0\pi$'', and ``$00\pi\pi$'' plane states, 
according to the stacking sequence.

\begin{figure}[h]
\begin{center}
\resizebox{\columnwidth}{!}{\includegraphics{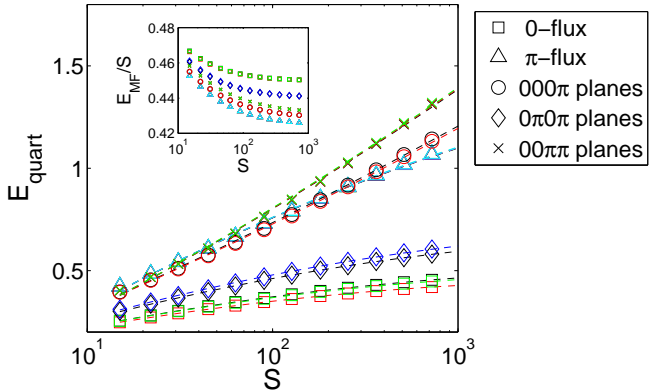}}
\end{center}
\caption{ \label{fig:scaling} \footnotesize (Color Online)
Quartic energy $E_\quart$ for $16$ classical collinear ground states.
$E_\quart(S)$ was obtained in the variational calculation.
The lines show a numerical quadratic fit in $\ln{S}$.
Each gauge family (represented by $2$-$6$ different states each)
is denoted by a different symbol, of which triangles denote
the harmonic ground states -- the $\pi$-flux states.
We show six $\pi$-flux states, and their energies are virtually indistinguishable
to the naked eye.
The total energy $E_\MF$ is shown in the inset. }
\end{figure}

We minimize the $E_\MF$ with respect to $\varepsilon$ at each value of $S$
and obtain the energy shown in the inset of Fig.~\ref{fig:scaling}.
We focused on the five simplest gauge families.
We minimize the $E_\MF$ with respect to $\varepsilon$ at each value of $S$
and obtain the energy shown in the inset of Fig.~\ref{fig:scaling}.
We focused on the five simplest gauge families.
We show the energies of all $16$ distinct Ising states 
belonging to the five gauge families.
Due to the exact invariance of the ($\varepsilon = 0$)
harmonic energy under the gaugelike
transformation, the total energies of states related by such transformations
are, as expected, indistinguishable in the inset, since the harmonic
term dominates.

In the main part of Fig.~\ref{fig:scaling} we show the anharmonic energy
$E_\quart$ for the same states.
As in the checkerboard lattice, the dominant part of the quartic energy
is quadratic in $\ln{S}$, and of the order $(\ln{S})^2$.
However, unlike the checkerboard lattice
(compare to Fig.~\ref{fig:check_equart}), we find that the energy
\emph{differences} between harmonically degenerate states are one to two orders
of magnitude smaller than the dominant quartic energy.

We first consider the dominant gauge invariant contribution to the 
quartic energy.
Since the invariants of the gaugelike transformation are products around loops,
we search for an effective Hamiltonian in terms of the fluxes $\Phi_{2n}$,
similar to the harmonic effective Hamiltonian~(\ref{eq:heff_harm}).
\begin{equation} \label{eq:heff_quart_gi}
E_\quart^\eff =A_0+ A_{6}(S) \Phi_{6} + A_{8}(S) \Phi_{8} +A_{10}(S) \Phi_{10}
+\cdots \,,
\end{equation}
where we find, numerically
\begin{eqnarray}
\label{eq:heff_quart_gi_coeffs}
A_0(S) &\approx&    0.300 + 0.0130 (\ln{S})^2  \,, \nonumber \\
A_6(S) &\approx&   -0.116 - 0.0030 (\ln{S})^2  \,, \nonumber \\
A_8(S) &\approx&   -0.022 + 0.0055(\ln{S})^2  \,, \nonumber \\ 
A_{10}(S)&\approx&  0.008 - 0.0021 (\ln{S})^2 \,. 
\end{eqnarray}
Note that for large $S$,
the signs of the coefficients $A_6$, $A_8$, and $A_{10}$
are opposite to $K_6$, $K_8$, and $K_{10}$ in the harmonic Hamiltonian.
The differences in signs among the $A_l(S)$ coefficients can explain why
some of the lines in Fig.~\ref{fig:scaling} appear to be convex and other 
concave: each family of states is dominated by different flux loop lengths $l$.

The gauge invariant terms can be heuristically explained in terms of the 
divergent modes:
the quartic energy is large for states that have a large number of divergent
modes.
%As was discussed in Sec.~\ref{sec:divergent}, the divergent modes are spanned by
%a set of real-space divergent modes,
%each with support on a subset of fcc sites $S_m$.
%In order for a divergent mode to exist on a given $S_m$,
It turns out~\cite{uh_thesis,uh_harmonic} that the number of divergent modes
is linearly related to the flux terms $\Phi_{2n}$: 
divergent modes proliferate
to the extent that the fluxes through loops of length $2n$ 
are $(-1)^n$.
%
%it turns out that any loop of length $n/2$
%(equivalent to diamond lattice or pyrochlore loops of length $n$),
%the flux $\varphi_\LL$ through the loop, defined in~(\ref{eq:flux}),
%must be $(-1)^{n/2}$.
%This ensures that the mock Ising model within $S_m$ is unfrustrated.
%Thus, the gauge invariant energy is largest when the product around hexagons is 
%negative and the product around loops of length eight in positive.
%Note that $\pi$-flux states do not have the largest quartic energy because
%the loops of length ten are all unsatisfied for these states.
%
%In fact, not only the divergent modes, but all ordinary modes of the bare
%Hamiltonian are gauge invariant.\cite{uh_thesis}
%In is only when one considers the details of our (variational) regularization
%scheme that the degeneracy is broken.

The above discussion of the gauge invariant quartic
energy~(\ref{eq:heff_quart_gi})
is somewhat moot, inasmuch as it is negligible compared
to the harmonic energy~(\ref{eq:heff_harm}), and it does not break the
gaugelike symmetry.
Nevertheless, one can clearly see in Fig.~\ref{fig:scaling}
that the anharmonic energy within
each gauge family is  not exactly the same, meaning that there is
a gauge-dependent term in the variational anharmonic energy.

\subsection{Gauge dependent terms and effective Hamiltonian} \label{sec:heff_gd}

Upon close inspection of Fig.~\ref{fig:scaling},
we see that some of the gauge families
have a larger dispersion in their quartic energies than others.
%Not surprisingly, given our knowledge of the checkerboard lattice
%states, these are the gauge families in which there is some anisotropy
%in the supported divergent modes.
%As in the checkerboard case, this leads to energy differences between 
%differently polarized tetrahedra, i.e., certain tetrahedra have a
%preferred polarization axis.
%Note that it is to be expected that, for the anisotropic families, the 
%energy differences are not quite as noticeable as in the checkerboard, 
%because there are many more divergent modes here and only fraction of them
%differ among the plane directions.
%[NOTE: All of this is not quite true for zero-flux states.]
%
But the quartic energy differences among
the ground states of the 
harmonic Hamiltonian --- the $\pi$-flux states ---
are much smaller than the gauge-invariant contribution.
We attribute this to the fact that, unlike the checkerboard lattice harmonic
ground states or even some pyrochlore gauge families, 
the $\pi$-flux states are completely uniform and isotropic
(at the gauge-invariant level),
and therefore
there is no reason for the harmonic degeneracy to be broken at the single-tetrahedron 
level (see the discussion of Sec.~\ref{sec:check_single}).
Indeed, in Appendix~\ref{app:ordinary} 
we show that,
the quartic energy due to {\it ordinary} modes of 
$\ham_\harm$ -- the dominant contribution --
is gauge invariant among $\pi$-flux states.
(This was not the case for the checkerboard case
of Sec.~\ref{sec:checker_select}.)
%%%%%%%%%%
We would expect any gauge-dependent terms
in an effective Hamiltonian to not be as local as those in, say,
Eq.~(\ref{eq:heff_check}).

\begin{figure}[h]
\begin{center}
\resizebox{\columnwidth}{!}{\includegraphics{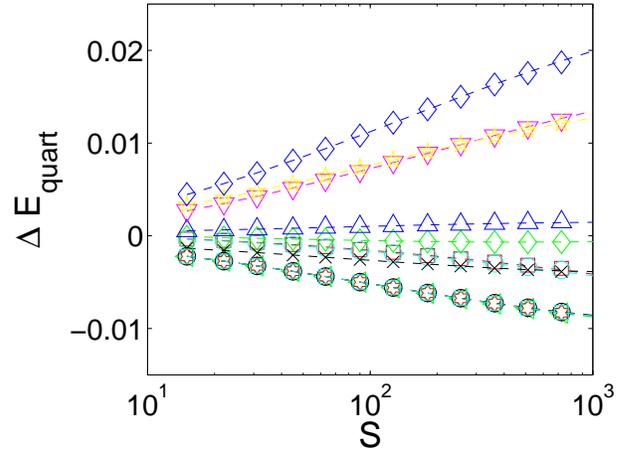}}
\end{center}
\caption{ \label{fig:delta_e} \footnotesize
Energy difference between $E_\quart$ of $12$ harmonic ground states
and the average of their energies $\overline{E}_\quart$.
By taking differences between energies, we eliminate the (dominant)
gauge-invariant term in the anharmonic energy.
Each dashed line shows a fit in $\ln{S}^2$, for one of the states..
Note that there are several overlapping symbols along the 
bottom line, representing the degenerate states described later
in the text (those with the maximum possible value of $\PP_6 = N_s/3$).}
\end{figure}

In Fig.~\ref{fig:delta_e}, we zoom in on the gauge dependent anharmonic energy,
by showing the difference
$\Delta E_\quart  \equiv  E_\quart  -   \overline{E}_\quart$,
where $E_\quart$ is calculated for $12$ $\pi$-flux
states, and
$\overline{E}_\quart$ is the mean quartic energy of the states shown in
the plot.

In order to systematically search for a ground state configuration
of the anharmonic effective Hamiltonian, 
we constructed
a large number of harmonic ground states using an algorithm for randomly
generating gaugelike transformations.\cite{uh_harmonic}
Within unit cells that we used, of up to $192$ sites, we believe that the 
algorithm performs an \emph{exhaustive} search for harmonic ground states.
About 350 states were found, inequivalent by lattice symmetries. (Notice
that non-cubic cells were tried; indeed, the optimal states described
below require a cell dimension that is a multiple of $3a/4$ in the 
stacking direction.)

The overall anharmonic energy (see Sec.~\ref{sec:scaling})
depends on $S$ as $(\ln S)^2$, as does its gauge-invariant part 
[see Eq.~(\ref{eq:heff_quart_gi_coeffs})]; is this also true for 
the gauge-dependent selection terms we seek?  From what has been
shown so far, that would be a plausible conjecture based on
the scaling of the total energy, as well as the checkerboard case.
Empirically, for each of our harmonic ground states, 
the $S$ dependence of its energy (including the gauge dependent part)
is well fitted by a linear or quadratic function $\ln{S}$ 
(as seen in Fig.~\ref{fig:delta_e}).
In fact, the checkerboard case is misleading: the anharmonic
selection there (unlike the pyrochlore) depends on the ordinary
spin-wave modes.  The analytic derivation in Sec.~\ref{sec:loop_expand}
shows the gauge-dependent term actually should scale as $\ln S/S$; 
we do not understand the discrepancy between this and the
numerical results.

\LATER{It appears to scale as 
$(\ln S)^2/S$?  But the $1/S$ factor is certainly inconsistent
with Fig.~\ref{fig:delta_e}).}

%%%%%%%%%%%%%
% For each of these states, the $S$ dependence of its energy can be extremely
% well fitted by a quadratic in $\ln{S}$ (as seen in Fig.~\ref{fig:delta_e}).
%  This is consistent with the view that 
% the decoupled quartic energy in Eq.~(\ref{eq:emf}) is a sum over
% products $\Gamma_{ij} \Gamma_{ji}$, with $\Gamma_{ij}$ linear in $\ln{S}$.
%%%%%%%%%%%%%
% We know both from numerics and from analytics (checkerboard) that
%     $\Gamma_{ij}$ scales as $\ln(\varepsilon)$. 
% Therefore (given that divergent part
%     of $\Gamma$ cancels out):
%            $E \sim A \varepsilon + B [\ln(\varepsilon)]^2/S^2.$
%        Minimizing with respect to $\varepsilon$ gives the relation to $\ln S$.
%%%%%%%%%%%%%%

\begin{figure}
\begin{center}
\resizebox{\columnwidth}{!}{\includegraphics{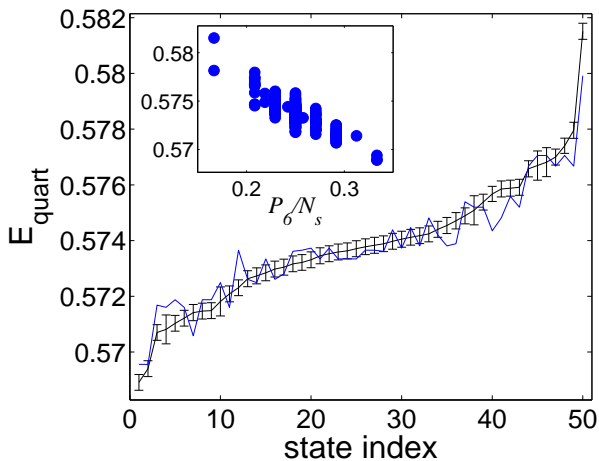}}
\end{center}
\caption{  \label{fig:eff_H} \footnotesize (Color Online)
The points with error bars are the numerical result
$E_\quart(S=100)$ for $50$ distinct
$\pi$-flux states, which had been found using our 
algorithm for generating gaugelike transformations.
(Note that these energies are monotonic by construction,
as the ``state index'' means simply the sequence when these 
energies are sorted.  Every seventh energy is plotted.
Shown for comparison are the energies predicted by the 
quartic effective Hamiltonian ~(\ref{eq:fit_gd}), using
best-fit values for the three coefficients.
The inset shows $E_\quart(S=100)$ as a function of the 
effective Hamiltonian's leading term, $\PP_6$.}
\end{figure}

In Fig.~\ref{fig:eff_H} we plot $E_\quart$ for the harmonic order ground
states at $S = 100$.
There are two sources of error in this calculation:
The first is 
the minimization error, represented by the error bars, which is due to the
difference in energy between consecutive value of $\varepsilon$ 
that we calculated, i.e. due to the ``grid'' in $\varepsilon$-space.
The second source of error 
is the grid used in integrating over the Brillouin zone, which is
equivalent to a finite (albeit large) system size.
This error becomes more significant for large values of $S$ (i.e., smaller
values of $\varepsilon$), where the singularity of the divergence lines becomes
narrower.
The results shown are for $15^3$ points in the Brillouin zone, for two different
magnetic unit cells: a cubic $128$ site unit cell, and a $96$ site tetragonal
unit cell.

As noted at the beginning of this section, 
we anticipate that an effective Hamiltonian should be 
represented by some sort of loop variables.
We now consider an effective Hamiltonian of the form
\begin{equation} \label{eq:fit_gd}
\Delta E_\quart^\eff = C_6(S) \PP_6 +C_8(S) \PP_8 +C_{10}(S) \PP_{10}
\,,
\end{equation}
where $\PP_l$ is equal to the number of loops of length $l$ composed solely of
satisfied AFM bonds.
The form (\ref{eq:fit_gd}) was partly inspired by
the effective Hamiltonian from Ref.~\onlinecite{uh_LN}, 
which is also a count of alternating loops (but with a 
broader definition of ``loop'' than here).
Eq.~(\ref{eq:fit_gd}) was guessed after fitting other forms with 
a variety of two- and four-spin terms involving the several closest
neighbors.  (Due to the ground-state constraint $\sum_{i\in \alpha} \eta_i=0$
and the $\pi$-flux constraint (\ref{eq:pi_flux}), there are numerous
linear dependencies among such terms.)

Also shown in Fig.~\ref{fig:eff_H} is a numerical fit to the
effective Hamiltonian ~(\ref{eq:fit_gd}).
For $S = 100$ we obtain
\begin{eqnarray}
\label{eq:C2n_100}
C_6 &=& -0.0621 \nonumber \,,\\
C_8 &=& -0.0223 \nonumber \,,\\
C_{10} &=& -0.0046 \,.  \nonumber \\
\end{eqnarray}
We ignore any constant terms here, as they belong in the gauge-invariant 
Hamiltonian~(\ref{eq:heff_quart_gi}).

While we cannot numerically repeat this calculation over a large range of
values of $S$, in order to find the functional dependence $C_l(S)$ with good
accuracy, we can obtain a rough fit by considering the small group of states
depicted in Fig.~\ref{fig:delta_e}. For these $12$ states we obtain
\begin{eqnarray} \label{eq:C2n_lnS}
C_6(S) &\approx& -0.015-0.004 (\ln{S})^2
\approx 0.05-0.03 \ln{S}
\,, \nonumber \\
C_8(S) &\approx&  0.002-0.002 (\ln{S})^2
\approx 0.04-0.02 \ln{S}
\,, \nonumber \\
C_{10}(S) &\approx& 0.0008-0.0005 (\ln{S})^2
\approx 0.009-0.004 \ln{S}
\,.
\end{eqnarray}
Over our range of $S=10$ to $1000$, either fit is
plausible but $\ln{S}$ is a litttle better than $(\ln{S})^2$.

\SAVE{Data are from Uzi's email 10/27/08.}

It must be noted that (at $S=100$) the coefficients in 
\eqr{eq:C2n_lnS} are bigger than \eqr{eq:C2n_100} by nearly
a factor of two; this is because the $12$ states used were 
not sufficiently representative.
Even though it is a rough fit, with significant error,
it is clear (see the inset in Fig.~\ref{fig:eff_H}) that for a large number
of states, the leading
order contribution to the energy is captured in Eq.~(\ref{eq:fit_gd}).
In particular, the numerical energy and the effective Hamiltonian
agree as to which states have the minimum and maximum energies.
As it turns out, this can be predicted from the first
term in (\ref{eq:fit_gd}):
the highest energy states are those with 
the highest $\PP_6$ value, namely $N_s/6$,
which means $1/6$ of all hexagons have alternating spin directions.
It can be shown that, for $\pi$-flux states, this is the smallest value that
$\PP_6$ can take.\cite{uh_thesis}.
The lowest energy states have $\PP_6 = N_s/3$ which is the highest 
possible value of $\PP_6$.

\subsection{Ground states}\label{sec:groundstates}

Since the $\PP_6$ term is largest, and in view of the results
just mentioned, it is a reasonable guess that the ground states 
are a subset of the ``hexagon-ground-states'' that maximize just
the $\PP_6$ term.  Since
(see Appendix \ref{app:octagons}) all hexagon-ground-states
are degenerate at the octagon term too, only the much weaker
10-loop term might split these states, this assumption
-- confirmed numerically in the results of subsection
\ref{sec:heff_gd} -- is very plausible.

\begin{figure}
 \begin{center}
 \resizebox{\columnwidth}{!}{\includegraphics{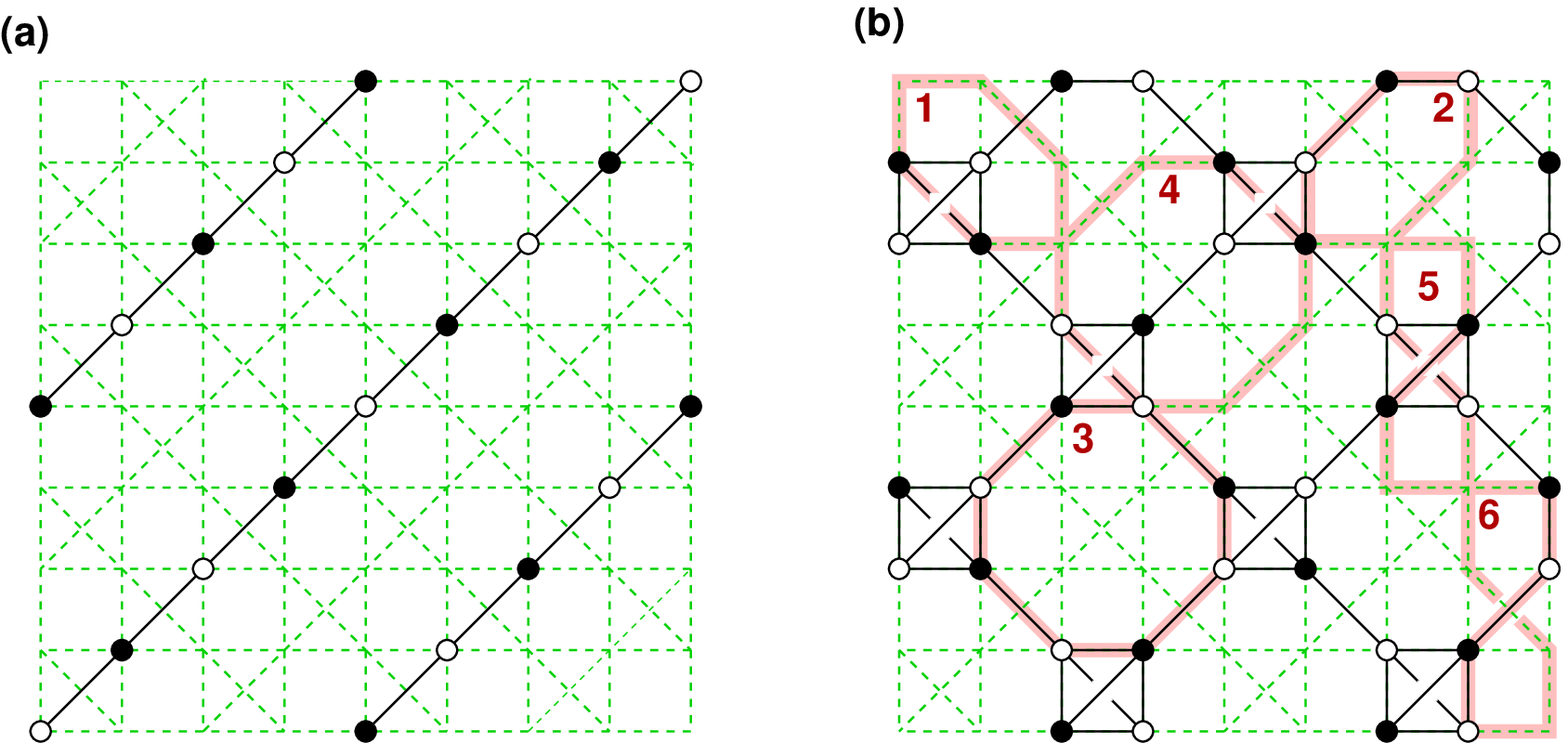}}
 \end{center}
 \caption{  \label{fig:slabs} \footnotesize (Color Online)
 Projection of the slabs which form the near-degenerate anharmonic 
 ground states of $\PP_6$, showing $A$ layer in (a) and $B$ layer in (b).
 The square shown is $2a \times 2a$.  Open and filled circles
  represent spin up and down.
\SAVE{This is consistent with Uzi's convention for up/down, 
as in Fig.~\ref{fig:checkerboard}}
%%%%%%%%%%%%%%%%%
\LATER{(Fall 2008-Jan 2009). It would be better to add panel 3 showing the 
superposition, and make the whole thing stretch across 2 columns.
But CLH doesn't remember exactly how that would work; would it 
be better to show (some) hexagons on this 3rd panel?  Deferred.}
%%%%%%%%%%%%%%%%%
Dashed lines are bonds outside the slab.
In (b), one loop is outlined (numbered) from each of the two classes 
of hexagon mentioned in text and in Table~\ref{tab:loops};
there are also four classes for octagon placement, numbered
3 -- 6 here.}
 \end{figure}

All hexagon ground states found 
could be constructed by layering two-dimensional slabs 
(see Fig.~\ref{fig:slabs});
they had unit cells of $48$ spins (or larger).
They were, within the numerical accuracy that we can obtain,
degenerate for all values of $S$. 
In fact, we found
these states share the same values of $\PP_l$ for all loop lengths 
that we calculated ($l \le   16$).  
Appendix~\ref{app:layers} explains these facts:
indeed, it is shown that 
all loops are identical for $l < 26$, and hence the stacked 
hexagon-ground-states must be \emph{exactly} degenerate up to that order,  
at least for any effective Hamiltonian written in terms of loops
[whether of the form \eqr{eq:fit_gd} or the form to be derived in
Sec.~\ref{sec:loop_expand}].

We conjecture that the stackings are, in fact, the only ground
hexagon-ground-states, but this is unproven since we have not
tried all possible unit cell shapes in the numerical enumeration.
Appendix~\ref{app:color-grstates} explains how one could
approach the ground state problem as a color-matching
problem, but does not solve it.

Although we shall find a different version of the effective Hamiltonian
in Sec.~\ref{sec:loop_expand}, this section is valid for that too.
All that matters is that the effective Hamiltonian depends on the
Ising configurations of loops, and that the hexagon term dominates.

\SAVE{This figure could be related to other ways of
depicting the layering shown in the thesis.}

% If we believe that the effective
% Hamiltonian of form~(\ref{eq:fit_gd}) is valid -- as confirmed 
% by the analytic derivation of Sec.~\ref{sec:loop_expand}, below --
% it is not surprising that the energy differences are tinier
% than our accuracy.

\section{Loop expansion} \label{sec:loop_expand}

In Sec.~\ref{sec:mft} we saw that in our self consistent theory, the 
mean-field Hamiltonian is proportional to the variational Hamiltonian
\begin{equation}
\ham_\MF = J^* \; \ham_\var
\end{equation}
In fact, it turns out that the quartic selection effects of $\ham_\MF$ can be
seen in the zero-point energy of $\ham_\var$, i.e. $J^*$ does not affect the 
selection.
Therefore, we can try to understand the origin of the quartic effective
Hamiltonian~(\ref{eq:fit_gd}), by studying $E_\var$, the zero point
energy of the variational Hamiltonian~(\ref{eq:variational}), 
treated as a purely harmonic problem.

In Refs.~\onlinecite{uh_harmonic} and \onlinecite{clh_harmonic} 
we developed an effective 
Hamiltonian for the harmonic zero-point energy by a real-space loop expansion.
Below (Sec.~\ref{sec:loop_var}),
we shall use the same method as motivation for Eq.~(\ref{eq:fit_gd}).
First, in Sec.~(\ref{sec:loop_harm}), we shall give a quick summary of the
results on $\ham_\harm$.
Next, we represent the variational Hamiltonian in similar matrix notation, and
repeat the loop expansion (for the leading order in $\varepsilon$), to derive an
analytic effective Hamiltonian (Sec.~\ref{sec:var_expand}).
In Sec.~\ref{sec:loop_compare} we discuss the obtained effective Hamiltonian and compare
it to the effective Hamiltonian we used in the numerical fit.

\subsection{Bare harmonic theory} \label{sec:loop_harm}

For this quick review of Ref.~\onlinecite{uh_harmonic},
it will be convenient to rewrite some
results of Sec.~\ref{sec:harmonic} using the matrix
notation of (\ref{eq:harm_mat}), as we note in each place.

The spin-wave modes in the unperturbed harmonic theory are 
the eigenvectors of the equation [equivalent to
(\ref{eq:eig})]
\begin{equation}
\etvec \Hvec \vvec_m =  \lambda_m \vvec_m\,,
\end{equation}
where $\Hvec$ can be written as
[equivalent to (\ref{eq:heis_mat})]
\begin{equation} \label{eq:h_w}
\Hvec = \half \Wvec^\dagger \Wvec\,.
\end{equation}
$\Wvec$ is a $N_s/2\!\times\!N_s$ matrix whose $(\alpha,i)$ element is $1$
if the pyrochlore site $i$ is in tetrahedron $\alpha$ and zero otherwise.

The spin-wave equation is transformed to the diamond lattice
(which is easier to deal with, since it has fewer loops), by defining
$\uvec_m \equiv \Wvec \vvec_m$. The diamond lattice modes satisfy the equation
[equivalent to (\ref{eq:ord_modes})]
\begin{equation}
\muvec \uvec_m =  \lambda_m \uvec_m\,,
\end{equation}
with the matrix $\muvec \equiv \half \Wvec \etvec \Wvec^\dagger$.

The elements of $\muvec$ only connect diamond-lattice nearest neighbors
and are equal to the value of $\eta$ at the center of the bonds.
\begin{equation}
\label{eq:muvec}
\mu_{\alpha \beta} = \left\{ \begin{array}{ll}
\eta_{i(\alpha \beta)} & \alpha\,,\beta\, \mbox{ nearest neighbors}\,,\\
0 & \mbox{otherwise}\,.
\end{array} \right.
\end{equation}
As before, $i(\alpha\beta)$ is the pyrochlore site at
the center of the diamond-bond $(\alpha \beta)$.
The zero point energy is $S \sum |\lambda_m|$, or in matrix notation
\begin{equation} \label{eq:trace}
E_\harm = S\tr{( \frac{1}{4} \muvec^2)^{1/2}} - S N_s\,.
\end{equation}
For each $\alpha$, the diagonal element $(\frac{1}{4} \mu^2)_{\alpha \alpha}$
 is equal to 
$1$, and thus the square-root can formally be Taylor-expanded 
in powers of $\muvec^2$ (or more exactly of $\muvec^2-4 \openone$).
\begin{equation} \label{eq:expanded}
E_\harm/S = 
  1+  \sum_{n=1} Q_{2n}
 \tr(\muvec^{2n})  -N_s\,,
\end{equation}
where the coefficients are
\begin{equation}
Q_{2n} \equiv   (-1)^{n+1} \frac{(2n-3)!!}{8^n n!}
\end{equation}
%%%%%%%%%%%%%%%%%%%%%%%%
\SAVE{On 1/18/09, 
def'n of $Q_{2n}$ was made consistent with harmonic paper 
(same factor of 2 in definition of $\muvec$).  In previous
drafts, a different convention was used (see Uzi email 10/25/08).}
%%%%%%%%%%%%%%%%%%%%%%%%

\SAVE{Notice $(2n-3)/2^n n! \sim n^{-3/2}$}
%%%%%%%%%%%%%%%%%%%%%%%%

The details of the expansion were given in Ref.~\onlinecite{uh_harmonic}, 
where the effective Hamiltonian 
(\ref{eq:heff_harm}), written in terms of $\{\eta_i\}$,
was derived from Eq.~(\ref{eq:expanded}).
However, the harmonic-order selection can be explained 
with a ``back-of-the-envelope''
argument, as in Ref.~\onlinecite{clh_harmonic}:
$\tr{\muvec^{2n}}$ is a sum of products of $\eta_{\alpha \beta}$ over
all closed paths in the diamond lattice.
Since any path that goes back and forth is independent of $\{\eta_i\}$, 
the only paths that contribute
non-trivial terms to the effective Hamiltonian are actual loops in the lattice. 
The first of these terms in for $n\!=\!3$ (corresponding to hexagons in the
pyrochlore lattice).
Thus, the first non-trivial term in the expansion favors states with negative 
hexagon fluxes  -- the $\pi$-flux states
with $\prod_{i \in \hexagon} \eta_i = -1$ [Eq.~(\ref{eq:pi_flux})].

\subsection{Variational Hamiltonian}\label{sec:loop_var}

The self-consistent theory (Sec.~\ref{sec:mft})
employs a variational Hamiltonian which has the same
form as the harmonic Hamiltonian but with $\Hvec$ replaced by
\begin{equation}
\Hvec_\var= \Hvec -\frac{1}{4} \varepsilon \etvec \Hvec \etvec +
\varepsilon \openone \,
\end{equation}
[to repeat (\ref{eq:variational}) and
(\ref{eq:hvar_all})].
Here $\varepsilon\!>\!0$ is the (small)
variational parameter.
The quartic energy is not equal to, but proportional to, the 
zero-point energy of the variational Hamiltonian [with 
its parameter $\varepsilon^*$ satisfying the self-consistency
equation  (\ref{eq:vareps-selfcons})].
Let us try to derive an expansion for this energy.

The spin-wave modes are eigenvectors of the equation
\begin{equation}
\lambda_m \vvec_m = \etvec \Hvec -\frac{1}{4} \varepsilon \Hvec \etvec+ \varepsilon \etvec
\vvec_m \,.
\end{equation}
Replacing $\Hvec$ by~(\ref{eq:h_w}), we obtain
\begin{equation} \label{eq:eom_var}
\lambda_m \vvec_m = \left( \half \etvec \Wvec^\dagger \Wvec -
\frac{1}{8} \varepsilon \Wvec^\dagger \Wvec \etvec +\varepsilon \etvec \right) \vvec_m\,.
\end{equation}
Clearly, the recipe for transposing this to the diamond lattice must
be generalized to a more complex form than before (which must
reduce to the old formulas in the case $\varepsilon=0$).
Luckily, thanks to the simple form adopted for our variational Hamiltonian 
(\ref{eq:variational})
it will suffice to expand the vector space of diamond modes from
one to two components.  Define the two vectors 

\begin{equation}
\uvec^1_m \equiv \Wvec \vvec_m \,,\qquad
\uvec^2_m \equiv \Wvec \etvec \vvec_m \,.
\end{equation}
For the case of $\varepsilon\!=\!0$, $\{\uvec^1_m\}$ corresponds to ordinary 
modes and $\{\uvec_m^2\}$ to generic zero modes.

It is convenient to introduce, analogous to $\muvec$, 
$\nuvec\equiv \Wvec \Wvec^\dagger$;
thus $\nuvec$ is
independent of $\{\eta_i\}$ and has nonzero elements
on the diagonal (with respect to the diamond-site index):
\begin{equation}
\label{eq:nuvec}
\nu_{\alpha \beta} = \left\{ \begin{array}{ll}
4 & \alpha=\beta \,, \\
1 & \alpha\,,\beta\, \mbox{ nearest neighbors}\,,\\
0 & \mbox{otherwise}\,.
\end{array} \right.
\end{equation}

Still defining $\muvec$ as in \eqr{eq:muvec},
we find [by multiplying Eq.~(\ref{eq:eom_var}) from the left
by $\Wvec$ and $\Wvec \etvec$]
the new equation of motion
\begin{equation}
\lambda_m
\left( \begin{array}{c} \uvec^1_m \\ \uvec^2_m \end{array}\right)
=  \Mvec
\left( \begin{array}{c} \uvec^1_m \\ \uvec^2_m \end{array}\right)\,.
\end{equation}
with the $2N_s\times 2N_s$ matrix $\Mvec$ defined as
\begin{equation}
\label{eq:Mvec}
\Mvec \equiv 
\left( \begin{array}{cc}
\muvec  & - \frac{1}{4}\varepsilon (\nuvec- 8 \openone) \\
\nuvec+ 2 \varepsilon \openone & - \frac{1}{4}\varepsilon \muvec 
\end{array}\right)  .
\end{equation}
The zero-point variational energy is
\begin{equation}\label{eq:trace-var}
E_\var = S \tr( \frac{1}{4} \Mvec^2)^{1/2} -S N_s\,.
\end{equation}
Note that now twice as many elements are summed in the trace
as were in the bare harmonic version (\ref{eq:trace}). 
One way to understand this is that the generic zero modes no longer 
have zero frequency and must explicitly appear in the
zero-point sum $S \sum |\lambda_m|$.

\subsection{Expansion of variational energy} \label{sec:var_expand}

The square root of (\ref{eq:trace-var})
can be formally expanded in exactly the sum 
Eq.~(\ref{eq:expanded}), but with the replacement
$\muvec^{2n} \to \Mvec^{2n}$.
In this trace expansion, each factor of $\muvec$
or $\nuvec$ hops us to a neighboring site --
with or without a factor of $\eta_i\eta_j$, respectively --
whereas a factor of $\openone$ does nothing.
We expect the lowest
order non-trivial terms in the expansion to be 
of order $6$ in $\muvec$, $\nuvec$, since it takes 
(at least) that many hops to complete a hexagon,
which is the smallest loop (in the pyrochlore lattice);
these contributions come  from the $+\! \tr(\Mvec^6)$ term

Furthermore, since $\varepsilon$ is a small parameter,
we shall expand the results in
orders of $\varepsilon$, keeping only the lowest order non-trivial term.
Notice that for every $\openone$ factor in \eqr{eq:Mvec}, 
we pay the price of one power of
$\varepsilon$ but do not gain a hop: hence, factors of $\openone$
cannot ever appear in a \emph{leading}  contribution.
Such factors serve to ``decorate'' a basic loop, so that 
the same contribution reappears coming from higher powers
of $\Mvec$ and of higher order in $\varepsilon$.
They play a role similar to (and in addition to) the
decorations by hops that retrace themselves, as 
found already in the bare harmonic theory~\cite{uh_harmonic}.

The upper-left block of $\Mvec$ corresponds to $\{\uvec^1_m\}$ -- the ordinary 
modes, whereas the lower-right block corresponds to $\{\uvec^2_m\}$ -- 
generic zero modes (that acquire nonzero frequency in the variational
Hamiltonian).
Since the matrix elements of the $\uvec^2$ sector always carry a factor 
$\varepsilon$, the leading order terms in the small-$\varepsilon$ expansion
will involve hops from the ordinary mode sector to the zero-mode sector 
and quickly return back. 
In this fashion, as conjectured in Appendix~\ref{app:ordinary},
we shall find explicitly that degeneracy breaking effects are 
due to the interaction
between generic zero modes and ordinary modes.

All nonzero terms in a trace represent paths 
$\WW$ of length $2l\leq 2n$ on the diamond lattice that start
and end on the same site (possibly retracing some bonds; 
also, $2n-2l$ is the number of factors $\nu_{\alpha\alpha}$
which are diagonal with respect to sites.
From here on we imagine having selected a particular path $\WW$,
which can be expressed 
as a sequence of pyrochlore sites (diamond-lattice bonds) 
$(i_1, i_2,\ldots,i_{2l})$; all terms in the traces
must be polynomials in the spins $\eta_{i_1}, \eta_{i_2}, ...$.
Then we considering the terms due to $\tr(\Mvec^{2n})$ 
at each order in $\varepsilon$.

The leading order [$\OO(1)$] terms involve only the upper-left block 
(ordinary modes) of $\Mvec$.
But it will be helpful to notice that $\tr(\muvec^{2n})= 
(2n) \varphi_\WW$, where $\varphi_\WW\equiv \prod _{j=1}^{2n} \eta_{i_j}$, 
%%% which is the product of $\eta_{i_j}$ along
%%% a path of length $2n-2$, as defined in 
which generalizes Eq.~(\ref{eq:flux}),
to a general closed path. 
(Here the factor $2n$ accounts for different cyclic permutations 
inside the trace, i.e. different places the same loop could have
been started.
Note that any retraced portions in $\WW$ have canceling 
contributions in the product $\varphi_\WW$.)
They are clearly gauge-invariant 
(See Appendix~\ref{app:gauge-correl})
by the definition of the gauge-symmetry 
as described in Sec.~\ref{sec:prior} and
are in fact exactly the same terms ($\muvec^{2n}$) that we had in
the bare harmonic theory [Eq.~(\ref{eq:expanded})].
Such terms in the effective Hamiltonian 
give the same value for all gauge-equivalent states,
so they do not split the harmonic-order degeneracy 
and are not of interest here.

In the next order,  $\OO(\varepsilon)$, we can have terms that take us 
out of the ordinary-mode sector in $M$ and into the zero-mode sector,
but come immediately back.
We obtain 
\begin{equation}\label{eq:ep1}
-\half S n Q_{2n}\varepsilon \tr\left[ \muvec^{2n-2}  (\nuvec - 8 \openone) \nuvec 
\right]\,,
\end{equation}
with the same $2n$ factor for cyclic permutations
The trace in Eq.~(\ref{eq:ep1}) contains two types of terms:
Firstly, taking the {\it site-diagonal} ($\alpha=\beta$) element
in each $\nuvec$, we obtain $4 \varphi_\WW$ (where
$|\WW|=2n-2$.) As noted above, this is gauge-invariant hence not
of interest.

Secondly, taking the site-{\it non-diagonal} 
elements of $\nuvec$, we obtain a products of all spins except
two adjacent ones, i.e.
\begin{equation}
\varphi_\WW \sum_j \eta_{i_j} \eta_{i_{j+1}}\,,
\end{equation}
where we adopted the notation convention $\eta_{i_{j+2n}}\equiv \eta_{i_j}$.
In (only) the special case of a $\pi$-flux state, all products $\varphi_\WW$ along
paths of the same topology are the same, and therefore a sum over {\it all}
paths of length $2n$ amounts
to a multiple of the classical energy $\sum_{\langle i j \rangle} \eta_i \eta_j$,
and does not split any degeneracies.
[More generally, within a family of non-$\pi$-flux states,
such terms do split the degeneracy and we must keep them.
This is probably the reason that the dispersion of 
quartic energies among non-$\pi$-flux states is notably
larger than in the $\pi$-flux or $0$-flux states
(see Fig.~\ref{fig:scaling}).]

Moving on to the terms of order $\varepsilon^2$, we have contributions
(i) from paths that hop once into the zero-mode sector (possibly
staying there for at most one hop) (ii) paths that hop twice into
the zero-mode sector (each time hopping back immediately):
\begin{eqnarray} \label{eq:ep2}
&&\frac{1}{8}S n Q_{2n} \varepsilon^2  \Big\{ 
\tr\left[ \muvec^{2n-3} (\nuvec - 8 \openone) \muvec \nuvec
-4  \muvec^{2n-2}  (\nuvec - 8 \openone)\right]
 \nonumber \\ &&
+  \frac{1}{2}\sum_{m=0}^{2n-4}
\tr\left[ \muvec^{m}  (\nuvec-8 \openone)\nuvec
\muvec^{2n-4-m} (\nuvec-8 \openone)\nuvec\right]
\Big\} \,. 
\end{eqnarray}
The prefactor of $1/2$ in front of the second trace corrects the
counting factor $2n$, since each placement of the pair of $\nuvec\nuvec$
factors is counted twice in the sum.

\SAVE{Each pair of $\nuvec$ factors involves one from the upper
right in \eqr{eq:Mvec}, and one from the lower left, for a net
factor $-\frac{1}{4}\varepsilon$, ignoring the $\openone$
in $\nuvec -4\openone$.  Also, any $\muvec$ between
the $\nuvec$'s gives a factor $-\frac{1}{4}\varepsilon$.  
Hence, we get the prefactor $(-1/4)^2 (2n)= (1/8)n$, QED.}

We now study Eq.~(\ref{eq:ep2}), seeking to  keep gauge dependent terms only.
%We must accept that all paths that go back and forth along the lattice (with no lattice loops), 
%must only contribute constants to the energy, since locally all Ising states that follow the tetrahedron
%%rule~(\ref{eq:classical_gs}) consist of the same paths 
%% [uzi old remark: references]
%Therefore, we will limit ourselves to loops in the lattice 
%% [uzi old remark: What about decorated loops?]
Start with the second term in the first trace,
inside the curly brackets:
$-4  \muvec^{2n-2}  (\nuvec - 8 \openone)$. 
In this term, only the site-diagonal elements in $\nuvec-8\openone$ can contribute, 
since the path has to be of an even length.
By the same arguments given above 
%%% before Eq.~(\ref{eq:ep1}), 
we just obtain $(-4)(4-8)\varphi_\WW$ which is gauge-invariant.
Next, the first term in the first trace
in Eq.~(\ref{eq:ep2}) produces one gauge invariant term 
(for diagonal elements of $\nuvec$)
plus one term that is gauge dependent:
\begin{equation}
\varphi_\WW 
\sum_{k=1}^{2n} \eta_{i_k} \eta_{i_{k+2}}
\equiv 
\varphi_\WW  T_\WW \, .
\end{equation}
Every factor inside the trace involves a hop to a  different  site.
Similarly, the sum over traces 
in~(\ref{eq:ep2}) results, for a path $\WW$, in terms
\begin{equation}
\varphi_\WW \!\! 
%%% \sum_{j,k: k \notin \{j-1,j,j+1\}}\!\! 
\frac{1}{2}  \sum_{j=1}^{2n} \sum _{k=j+2}^{2n+j-2}
\eta_{i_j} \eta_{i_{j+1}} \eta_{i_k} \eta_{i_{k+1}} 
\,,
\end{equation}
plus gauge-invariant terms that result from diagonal elements in $\nuvec-4\openone$.
This can be simplified into 
$ \frac{1}{2}\varphi_\WW (U_\WW^2 - 2 T_\WW)$, 
where we define
\begin{equation}
U_\WW \equiv \sum_{k=1}^{2n} \eta_{i_k} \eta_{i_{k+1}}\,.
\end{equation}
%%%
Merging these two expressions together, we obtain, up to gauge invariant terms:
\begin{equation} \label{eq:ep2_U}
\frac{1}{16}S n Q_{2n} \varepsilon^2  
\sum_{|\WW|=2n} \varphi_\WW U_\WW^2  \,,
\end{equation}

It is easy to see that only actual loops contribute interesting terms to Eq.~(\ref{eq:ep2_U}) -- 
all paths that go back and forth along the lattice add up to terms that are equal for all states
that obey the ``tetrahedron rule'' $\sum_{i\in\alpha} \eta_i =0$.
Thus the anharmonic energy, to order $\varepsilon$, can be expressed as a sum over lattice loops
$\{\LL\}$
\SAVE{Coefficient here is 1/2 as big as in 
earlier drafts due to the factor from overcounting.}

\begin{equation} \label{eq:loop_evar}
E_\var(\mathrm{gauge\ dep.}) =  
\frac{\varepsilon^2 S}{16}  \sum_{n=3} n  \tilde{Q}_{2n}
\!\!\! \sum_{|\LL|=2n} \!\!\!  \varphi_{\LL} |U_\LL|^2 + \OO(\varepsilon^3) \,.
\end{equation}
Here, the coefficient $\tilde{Q}_{2n}$ is not quite the same as $Q_{2n}$, since loop
terms of length $2n$ are renormalized by ``decorated loops'' of longer lengths.
These are paths that go along the loop with additional back-and-forth paths added to them.
Such decorated loops have been discussed extensively, for related problems,
in Refs.~\onlinecite{uh_LN,uh_harmonic},
and can be summed up by use of simple combinatorics.

Eq.~(\ref{eq:loop_evar}) is the final result of this section
and defines the quartic effective Hamiltonian $E_\quart^\eff$.
Assuming we chose $\varepsilon=\varepsilon^*(S)$, 
the self-consistent value, then each term in
$E_\quart^\eff$ is $\propto S \varepsilon^2$,
i.e.  $\propto (\ln S)^2/S$, in light of (\ref{eq:vareps-S}).
We do not understand the discrepancy (by a factor of $1/S$)
with with logarithmic scaling of the fitted effective 
Hamiltonian in Fig.~\ref{fig:delta_e} and Eq.~\eqr{eq:C2n_lnS}.

\subsection{Discussion of loop derivation} \label{sec:loop_compare}

With ~(\ref{eq:loop_evar}) we can completely understand 
the essential features of the quartic effective Hamiltonian, 
and how the analytic results of Sec.~\ref{sec:loop_expand}
relate to the (prior) fit results of Sec.~\ref{sec:num}.
Eqs.~(\ref{eq:loop_evar}) and ~(\ref{eq:fit_gd}) are both sums
over the same kinds of loops. 
The terms do {\it not} have the same analytic functional form,
but are related, in being minimized by the same configuration of
alternating spins around that loop.
Hence we understand how \eqr{eq:loop_evar} and ~(\ref{eq:fit_gd}) 
tend to be optimized by the same
configurations, and hence why ~(\ref{eq:fit_gd}) was a good
approximation of the correct effective Hamiltonian.

First, the leading order term in~(\ref{eq:loop_evar}) is due to hexagons.
Since the number of AFM bonds within a single hexagon (in a $\pi$-flux state) 
can be $2$, $4$, or $6$, 
and since $|U_\LL|=2$ is the same for both the case of $2$ AFM bonds and the case of $4$ AFM bonds, then
\begin{equation}
\label{eq:hexagon-UL}
\sum_{\hexagon} \varphi_\LL |U_\LL|^2 = -32 \PP_6 + \mathrm{const}\,.
\end{equation}
Thus, this term is in exact agreement the leading term in
with Eq.~(\ref{eq:fit_gd}). 
It accounts for the largest contribution, sufficiently
large that our ground state search can be limited
to the subset optimizing the hexagon term minimizing $|U_6|^2$ or equivalently
maximizing $\PP_6$.

The next to leading term is due to octagon loops.  Already at this order,
$|U_8|^2$ is {\it not} independent of $\PP_8$.
But, within $\pi$-flux states, an octagon has $\varphi_\LL\!=\!+1$, 
and since $Q_8\!=\!-1$, then a large $|U_\LL|$ is favored. 
Clearly, a large $\sum_{\small{\octagon}} |U_\LL|^2$  means
a tendency to alternate and this correlates with large $\PP_8$,
meaning that a large $\PP_8$ is favored by Eq.~(\ref{eq:loop_evar}).
(In any case, among states optimizing
\eqr{eq:hexagon-UL}, the octagon terms are always
the same: see Appendix~\ref{app:octagons}.)

As for loops of length $10$ or longer, the situation is further complicated 
because the pyrochlore lattice has more than one kind (modulo symmetries) 
and $\varphi_\LL$ may not be the same 
for different kinds of loop.
Indeed, one kind of $10$-loop has $\varphi_\LL\!=\!+1$ 
while another kind has $\varphi_\LL\!=\!-1$, in $\pi$-flux states, 
Therefore some of the $10$-loops actually prefer to have a small 
$|U_\LL|$, and it is not certain {\it a priori} that 
$\PP_{10}$ should be maximized.

But the role of larger loops simplifies in the 
special case of the hexagon-ground-states 
(the subset of $\pi$-flux states that optimizes $\PP_6$).
The octagon terms (of either the fitted effective Hamiltonian
~\eqr{eq:fit_gd} or the analytic one~\eqr{eq:loop_evar})
turn out to be the same for any of these states.
Furthermore, at least for the stacked hexagon-ground-states 
found by the exhaustive search in Sec.~\ref{sec:num},
and described in Sec.~eref{sec:groundstates}, 
many more terms are degenerate too.
Each term appearing in Eq.~(\ref{eq:loop_evar}) is the same in {\it every} 
state of this family, at least up to the terms for $|\LL|\!=\!16$.
Thus the degeneracy is broken only from a quite long loop
that we anticipate to have a minuscule coefficient.

\section{\label{sec:concl}Discussion}

We have calculated the anharmonic corrections to the spin-wave energy in 
the pyrochlore, and found that they break the degeneracy between the
various harmonic ground states.
We managed to numerically construct an effective Hamiltonian, and in
Sec.~\ref{sec:loop_expand}, obtained an understanding of its terms.

In retrospect, we should not have been surprised to find
that the effective Hamiltonian is
written in terms of loop variables. After all, in any collinear configuration,
the local environment that each spin sees is the same for all sites.
If the centers of the simplexes were put on a Bethe lattice rather than
a diamond lattice, then all collinear configurations would be related by lattice 
symmetries and would therefore have the same energy (as was found explicitly
in the harmonic theory of Ref.~\onlinecite{uh_harmonic} and the large-$N$
theory of Ref.~\onlinecite{uh_LN}, and in analogy to
Ref.~\onlinecite{doucot}).
Thus any degeneracy-breaking terms \emph{must} arise from lattice loops, so it
is plausible that the effective Hamiltonian could be written explicitly in terms
of loop configurations, but there are still multiple possibilities:
the analytic derivation said the loop term is the square of the number of
antiferromagnetic bonds along it [Eq.~\eqr{eq:loop_evar}]
whereas a good numerical fit was obtained to a Hamiltonian that 
counts only the loops with {\it all} bonds antiferromagnetic
[Eq.~\eqr{eq:fit_gd}].

The anharmonic Hamiltonian is dominated by the smallest loops, the
``hexagon'' terms.  The hexagon term's ground states are degenerate,
having an $\OO(L)$ entropy; we conjectured that the stacked family
in Sec.~\ref{sec:groundstates} are {\it all} of its ground states, 
but we did not demonstrate it (see Appendix \ref{app:color-grstates}).
Within those states at least (and certainly to octagon order in any
hexagon-ground-state), the count of many longer loops is constrained
so that only a tiny term can break the degeneracy, which (for the
stacked family at least) is only at the length 26 loops.  To the accuracy layers
of our numerics, all the stacked ground states are degenerate.

What do our results say for realistic spins?
First of all, the ``small parameter'' turned out to be $1/\ln S$, 
which is not really small except at unphysical spin lengths
[$S=10$--$10^3$ were used for numerical fits in Sec.~\ref{sec:heff_gd}].
Still, our argument that only loop terms can break degeneracies
still applies, so we expect the effective Hamiltonian takes
similar functional forms for realistic $S$.  It appears that
only the first (hexagon) loop term will be important, since 
this will fix the values of the next few terms and only some
very long loops will cause quite small splittings in these
energies.  So in practice this leaves
a massive but non extensive degeneracy $\exp(O(L))$, as was already 
the case for the harmonic ground state~\cite{uh_harmonic}
(but with a smaller coefficient of $L$).

It is worth noting that the anharmonic selection effects in the pyrochlore 
turn out to be much weaker than in other closely related lattices:
the two-dimensional checkerboard and kagom\'e lattices.
%%%
In the checkerboard lattice, which we discussed in Sec.~\ref{sec:checker},
many of the details are the same as in the pyrochlore:
it is composed of corner sharing tetrahedra, the spin-wave Hamiltonian 
is the same, and the harmonic ground states are collinear states 
with uniform fluxes.
Nevertheless, because of the anisotropy inherent to the two-dimensional
checkerboard, the anharmonic energy breaks the harmonic degeneracy
at the lowest order terms, of order $(\ln{S})^2$.

In the kagom\'e lattice, the anharmonic selection is even stronger:
first, there are cubic (in spin $\sigma^{x/y}$) anharmonic spin-waves terms.
In addition, because of the anisotropy between in-plane and out-of-plane 
fluctuations about the coplanar states,
\emph{all} harmonic zero modes possess divergent 
fluctuations and therefore the anharmonic energy scales as a power law in
$S$.\cite{chubukov,chan,chan_thesis}

Finally, we would like to mention that a similar calculation can be carried out
in the case of collinear states with nonzero magnetization, in the presence
of a magnetic field. Such magnetization plateaus have been the subject of 
numerous recent studies.\cite{ueda,penc,bergman,zhp_field,hassan}
Our own harmonic work on the subject concluded that for 
a magnetic field that induces a collinear spin arrangement such that
$\sum\eta_i=2$ in each tetrahedron, the degenerate harmonic ground states are
zero-flux states.\cite{uh_harmonic}
One could develop a self consistent variational treatment analogous to the one
in this paper, to find that quartic ground state. Due to the asymmetry between
$\uparrow$ spins and $\downarrow$ spins, there will be two independent
variational parameters.
In particular, the bond variables $\Gamma_{ij}$ are no longer expected to 
satisfy Eq.~(\ref{eq:gamma_emp}). Rather, we expect the dominant terms in
$\Gamma_{ij}$ to be $\Gamma^{0}+(\eta_i-\eta_j) \Gamma^{(1)}+
\eta_i \eta_j \Gamma^{(2)}$ (see Appendix~\ref{app:ordinary}).

\begin{acknowledgments}

This work was supported by the NSF, under grant DMR-0552461.
We acknowledge the Cornell Center for Materials Research for use of its
computer resources.

\end{acknowledgments}

\appendix

\section{Ordinary modes} \label{app:ordinary}

To attempt to understand the results of the anharmonic calculation,
the first thing we try is to calculate the contribution
to the anharmonic energy due to ordinary modes, as we did,
for the checkerboard lattice, in Sec.~\ref{sec:checker_select}.
The reason that we focus on ordinary modes is that, unlike generic zero-modes,
we know how they transform under gaugelike transformation.
In the checkerboard case, we saw (Sec.~\ref{sec:checker_select}) that
the anharmonic selection can be explained in terms of 
the correlations due to ordinary modes 
in the harmonic Hamiltonian.
As we shall see below, this is not true for the pyrochlore lattice, i.e.
the ordinary modes produce a gauge-invariant quartic energy.

\subsection{Calculating correlations}
\label{app:ord-calc}

An ordinary mode $\vvec_m$ is a mode that can be expressed in terms of
a diamond-lattice mode  $\uvec_m$ by Eq.~(\ref{eq:ord_modes}).
The correlation function $G_{ij}$ was shown in Sec.~\ref{sec:diag} to be written
as a sum over the spin-wave modes
\begin{equation} \label{eq:fluct_2}
G_{ij} = \sum_m \frac{S}{2|\vvec_m^\dagger \etvec \vvec_m|}
v_m(i) v_m(j) \,.
\end{equation}
%
%%%%%%%%%%%%%%%%%%%%
Restricting ourselves to the contribution of ordinary modes 
(denoted henceforth by superscript ``$\mathrm ord$''), 
and using Eqs.~(\ref{eq:fluct}) and~(\ref{eq:eta-norm-ord}),
\begin{eqnarray}\label{eq:Gord-uu}
G^\ord_{ij} &=& {\sum_m}^\ord \frac{S}{2|\lambda_m|}
\eta_i \eta_j \sum _{\alpha, \beta:i \in \alpha, j\in \beta}
u_m(\alpha) u_m(\beta) \nonumber \\ 
&=& \eta_i \eta_j \sum _{\alpha, \beta: \in \alpha, j\in \beta}
\! \! g_{\alpha \beta}\,.
\end{eqnarray}

For (\ref{eq:Gord-uu}) we defined, 
in analogy with~(\ref{eq:fluct_2})
\begin{equation} \label{eq:g_ab}
%\langle \zeta_\alpha \zeta_\beta \rangle \equiv
g_{\alpha \beta} \equiv
{\sum_m}^\ord \frac{S}{2|\lambda_m|}
u_m(\alpha) u_m(\beta) \,.
\end{equation}
%% ``Note that $\sum_{\alpha\beta: i\in\alpha, j\in\beta} 
%% \langle \zeta\alpha \zeta_\beta \rangle$ consists of four terms;
%% on the diamond lattice, one is on-site ($\alpha=\beta$), two
%% are nearest-neighbor, and one is second-neighbor.''}

We need the bond variables ~(\ref{eq:corrdiff}), 
for a nearest-neighbor pair $(ij)$, since that is how 
correlations enter our results [such as (\ref{eq:hmf_gamma})].
To express this for a particular pair, let $\alpha$ be the 
common diamond site, and let $\beta$ and $\beta'$ be the
diamond sites at the far ends of the bonds on which sites 
$i$ and $j$ sit, respectively. Then
\begin{eqnarray} 
\Gamma^\ord_{ij} &&=
g_{\beta \beta} + g_{\beta \alpha} - g_{\beta' \alpha} - g_{\beta \beta'} \,.
%\langle \sigma_{\beta}^2 \rangle +
%\langle  (\sigma_{\beta} -\sigma_{\beta'}) \sigma_{\alpha} \rangle-
%\langle  \sigma_{\beta} \sigma_{\beta'} \rangle\,.
\label{eq:gamma_prime}
\end{eqnarray}
Note that the last line consists of one on-(diamond)-site correlation 
function, (the difference of) two nearest neighbor correlations, and one
second-neighbor diamond mode correlation.

\subsection{Using the gaugelike symmetry}\label{app:gauge-correl}

Although we have been considering one particular classical configuration, we 
can make use of the concept of gaugelike transformations (discussed in
Sec.~\ref{sec:ordinary}). The important points are the following:\\
(i) Under a gaugelike transformation $\boldsymbol{\tau}$
(recall $\tau_\alpha \!=\! \pm1$)
the diamond-lattice spin-wave modes transform
$u_m(\alpha) \to \tau_\alpha u_m(\alpha)$;
$\eta_{i(\alpha\beta)} \to \tau_{\alpha} \tau_{\beta} \eta_{i(\alpha\beta)}$.\\
(ii) If two states have the same products of $\{\eta_i\}$ (flux)
around each loop in the lattice, they are related by a gaugelike transformation.\\
(iii) In particular, if the state has a uniform flux arrangement,
(e.g. the $\pi$-flux states), then {\it any} new  configuration
generated by a lattice-symmetry operation can 
alternatively be generated by a gaugelike transformation.\\
The consequences of these points is that, for the $\pi$-flux states
\begin{equation}
\Gamma^{(0)} \equiv g_{\alpha \alpha}
\qquad \mbox{is independent of $\alpha$,}
\end{equation}
(since a gaugelike transformation would take $\alpha$ to $\beta$ for any two
diamond-sites $\alpha$ and $\beta$).
Similarly, it is easy to find that for nearest neighbor (diamond) sites  $\alpha$, $\beta$
(sharing site $i$):
\begin{equation}
\label{eq:Gamma1_indep}
 \Gamma^{(1)} \equiv \eta_i g_{\alpha \beta}
, \qquad \mbox{independent of $i$,}
\end{equation}
and for next-nearest-neighbor (diamond) sites $\beta$, $\beta'$, connected by bond $(ij)$:
\begin{equation}
\label{eq:Gamma2_indep}
%\langle \sigma_{\alpha} \sigma_{\beta}\rangle
 \Gamma^{(2)}\equiv - \eta_i \eta_j g_{\beta \beta'}
,\qquad \mbox{independent of $(ij)$} .
\end{equation}
In (\ref{eq:Gamma2_indep}), 
the sign was set so that $\Gamma^{(2)}$ would be positive.
Plugging these into~(\ref{eq:gamma_prime}), we obtain
\begin{equation}
\Gamma^\ord_{ij}=\Gamma^{(0)}+ (\eta_i-\eta_j) \Gamma^{(1)}+
\eta_i \eta_j \Gamma^{(2)}\,.
\end{equation}
Since $\Gamma^\ord_{ij}$ must be invariant 
under a global spin-flip, we must have
$\Gamma^{(1)}\equiv0$ and we obtain
\begin{equation} \label{eq:gauge_emp}
\Gamma^\ord_{ij}= \Gamma^{(0)}+\eta_i \eta_j \Gamma^{(2)}\,.
\end{equation}
Eq.~(\ref{eq:gauge_emp}) is the key result of this appendix, 
the justification of Eq.~(\ref{eq:gamma_emp}).
It should be noted that $\Gamma^{(0)}$ and $\Gamma^{(2)}$ are both
infinite in the bare harmonic theory, and are regularized by the variational
scheme.
Here we assume that the regularization would not change the fact that
$\Gamma^{(0)}$ and $\Gamma^{(2)}$ are spatially invariant and gauge-independent.

Furthermore, by the argument above, $\Gamma^{(0)}$ and $\Gamma^{(2)}$ are the
same for any harmonic ground state ($\pi$-flux state).
Inserting Eq.~(\ref{eq:gauge_emp}) into the mean-field energy~(\ref{eq:emf}),
we quickly find that the ordinary modes'
contribution to the anharmonic energy is gauge-invariant:
\begin{eqnarray}
{E_\MF}^\ord &=& -\sum_{\langle ij \rangle} \eta_i \eta_j 
\left( \Gamma_{ij} + \Gamma_{ji}-\frac{1}{S^2} \Gamma_{ij} \Gamma_{ji} \right)
 \\ &=&
-\sum_{\langle ij \rangle} 
\left[  \left( 2\Gamma^{(0)} -\frac{(\Gamma^{(0)})^2 +(\Gamma^{(2)})^2}{S^2}
\right)
\eta_i \eta_j \right.
\nonumber \\ && \left.
+ 2 \left( \Gamma^{(2)} - \frac{(\Gamma^{(0)} \Gamma^{(2)}}{S^2}\right)\right]
\nonumber \\ = N_s
 &&\!\!\!\! \!\!\!\!\left[ 2\Big(\Gamma^{(0)}-3\Gamma^{(2)}\Big) -
\frac{(\Gamma^{(0)})^2 +(\Gamma^{(2)})^2-6\Gamma^{(0)} \Gamma^{(2)}}{S^2}\right]
\nonumber
\end{eqnarray}
Note that the arguments above do not apply to the checkerboard lattice, where
all bonds are \emph{not} equivalent by gauge-transformations -- there is no transformation that can take 
a diagonal bond and turn it into a horizontal or vertical bond. Therefore, the
correlations calculated from ordinary modes are sufficient to break the harmonic-order degeneracy in that case
, as we find in Sec.~\ref{sec:checker}.

\subsection{Relation of $\Gamma^{(0)}$ to $\Gamma^{(2)}$}

We take a moment to note that the parameters 
$\Gamma^{(0)}$ and $\Gamma^{(2)}$ are not independent.
We start from the variational Hamiltonian (Sec.~\ref{sec:ham-var}).
Notice that $\langle \ham_\var \rangle  =  E_\harm + \OO(\varepsilon)$,
On the one hand, 
$\langle \ham_\var \rangle  =  E_\harm + \OO(\varepsilon)$,
since [look at (\ref{eq:variational})] we could always do this
well by using the wavefunction of the bare harmonic $\ham_\harm$.
On the other hand, (\ref{eq:hamexp_Gamma}) 
[which is part of the expectation (\ref{eq:emf})] 
contains terms in $\{ \Gamma _{ij} \}$ which
are divergent as $\varepsilon\to 0$:
these must cancel out, at the dominant order.
In other words, 
$\Gamma_{ij}+\Gamma_{ji}$, must cancel out.
\begin{eqnarray}\label{eq:ham_var_dominant}
\langle \ham_\var\rangle_{\rm dominant} &=&
\sum_{\langle ij \rangle} \eta_i \eta_j (\Gamma_{ij} + \Gamma_{ji}) 
\approx N_{\rm FM} \big[ \Gamma^{(0)} + \Gamma^{(2)} \big] \nonumber \\
&+& N_{\rm AFM} \big[ \Gamma^{(0)} - \Gamma^{(2)} \big]
= \OO(\varepsilon)\,.
\end{eqnarray}
Since (\ref{eq:gamma_emp})  says
$\Gamma_{ij}$ (at dominant order) just depends on the sign of $\eta_i \eta_j$, 
the sum groups into $N_{\rm FM}$ terms for the FM bonds
and $N_{\rm AFM}$ terms for the AFM bonds.  But since 
$N_{\rm AFM} = 2 N_{\rm FM}$  in any ground state, 
\begin{equation}
\label{eq:gamma1to3}
\Gamma^{(2)}(\varepsilon) / \Gamma^{(0)}(\varepsilon) \to \frac{1}{3}\,,
\end{equation}
valid for the limit $\varepsilon \to 0$.
Numerically, 
$\Gamma^{(2)}$ appeared to be between $\Gamma^{(0)}/3$ and  $\Gamma^{(0)}/2$,

\LATER{Equation \eqr{eq:gamma1to3} isn't, in fact, used for any
other big purpose.}

\subsection{Role of generic zero modes}
\label{sec:app-genericzero}

Note that in the entire discussion, we have ignored the generic zero modes.
Recall that divergent modes occur along lines in the Brillouin zone at $\qvec$
values for which the ordinary modes' frequency goes to zero.
For $\qvec$ values close to these divergence lines,
the zero-modes and small-frequency
ordinary modes become close to each other (until they merge on the divergence
lines; divergent modes are both ordinary and zero modes).
The nearly divergent generic zero modes' contribution to the correlations 
mirrors the contribution of the nearly divergent ordinary modes,
and therefore $\Gamma_{ij}\approx 2 \Gamma^\ord_{ij}$ and it has the
same functional form~(\ref{eq:gauge_emp}). 

In the self-consistent variational theory, the generic zero modes and the
ordinary modes in the vicinity of the divergent lines interact strongly and, 
in fact, this interaction is responsible for the degeneracy-breaking, as we observe
in Sec.~\ref{sec:loop_expand}.

\SAVE{In light of Sec. VI, 
the $\Delta \Gamma_{ij}$ is due to interaction with the ordinary zero-modes
(which we found in the analytics to break the degeneracy).}

\section{Stacked ground states} \label{app:layers}

In this appendix, we analyze analytically the ground states of the 
effective Hamiltonians found in Sec.~\ref{sec:heff_gd} and
Sec. \ref{sec:var_expand}, as summarized in 
Sec.~\ref{sec:groundstates}.  We assume a stacked spin configuration
(see Fig.~\ref{fig:slabs})
as this is what emerged from numerics; however, this is not yet proven. 

\subsection{Layer stackings}

The pyrochlore sites can be broken into 
a stack of layers, each $a/4$ thick, where
$a$ is the lattice constant of the conventional cubic cell.
The hexagon-ground-states are stackings of two kinds of 
slabs parallel to (say) the $(001)$ plane: 
thin ``$A$'' slabs (thickness $a/4$) and thick ``$B$'' slabs 
(thickness $a/2$), which are stacked alternating $A$ and $B$.
A thin slab has one level of chains along the $[110]$ or $([{\bar 1} 0]$
direction,  along which the spins repeat the pattern 
``$+-+-$'.
%%% $\uparrow\downarrow\uparrow\downarrow\ $;
This pattern is reversed under a shift of $[a00]$ or [$0a0]$, 
so the periodicity is $\sqrt{2}a \times \sqrt{2} a$ 
within a thin slab.

A thick slab has two layers of spins, which form chains along
the $[110]$ and $[1\bar 1 0]$ directions, 
repeating the pattern 
``$++--$'', 
%%% $\uparrow\uparrow\downarrow\downarrow$, 
such that the chain spins are
parallel and the interlevel bonds are AFM in every tetrahedron
spanning those two layers; 
within the thick slab, the spin pattern has a period $2a\times 2a$.

The inter-slab spin couplings cancel, so each 
slab has an independent choice of two ways to align its spins.
When there are $m$ slabs of either kind, for a linear dimension 
in the stacking direction $L_z= m (3/4)a$, 
the number of stacked spin states is thus $3\times 2^{2m}=3 \times 2^{8L_z/3}$.
This includes three possible possible offsets (by multiples of $a/4$)
in the $z$ direction for the start of the stacking.
[In a rectangular cell where $L_x$ or $L_y$ are also multiples of $3a$
(see below), we add similar terms counting possible spin stackings 
in the $x$ or $y$ directions.]
Notice, apart that initial offset, the actual sites forming the
layers are determined; only the spin directions are free.

As a side remark, we can compare this to the family of harmonic 
ground states for the pyrochlore as described in 
Ref.~\onlinecite{clh_harmonic}:
that was a stacking of only $A$ slabs.  
The family of ground states of the 
effective Hamiltonian derived in the large-$N$ theory
for the pyrochlore~\cite{uh_LN}
is a stacking of alternating thin $A'$ and $B$ layers. 
The $A'$ slab differs from the $A$ layer shown in
Fig.~\ref{fig:slabs}(a) in that the spin patten is the
{\it same} under a shift of $[a00]$.  
%%%%%%%%%%
\footnote{These states were illustrated in Fig.~3 of Ref.~\onlinecite{uh_LN}
with the stacking in the $x$ direction, the $B$ layers being those
with all bonds AFM, or gray in that figure.}

\SAVE{
In light of the last observation, the large-$N$ stacking also
can repeat only after 12 spin layers.  So how come the degeneracy
can be broken there by loops of $l=16$?}

Now we examine the slab stacking more carefully.
The way a $B$ layer adjoins $A$
layers on opposite sides forces successive $A$ layers to 
have opposite orientations: i.e., if one slab has chains along
$[110]$ the next one has them along $[1\bar1 0]$, etc.
On the other hand, the way an $A$ layer adjoins its neighboring
$B$ layers requires these $B$ layers to have a relative shift in
the $xy$ plane of $(a/4)[110]$ or
$(a/4)[1\bar 10]$ parallel to the $A$ layer's chains.
Hence, the $xy$ offset of the $B$ layer cycles through
all four possible values in successive $B$ slabs.
The result is any periodic stack must have $m$ even,
e.g. $m=2$ has a period $[a/2, a/2, 3a/2]$ producing
centered tetragonal cell.  To directly repeat the
same layer requires $m$ to  be a multiple of four, so
the shortest cell ($m=4, L_z=3a)$ contains 12 layers of sites.

\subsection{Counting short loops}

%TTTTTTTTTTTTTTTTTTTTTTTTTTTTTTTTTTTTTTTTTTTTTTTTTTTTTTTTTTTTTTTTTTTTTT
\begin{table}
\caption{Types of spin patterns in $\pi$-flux hexagon loops.
Only hexagons with a loop product $\varphi_\LL=-1$ are included.
Values are given for the two effective Hamiltonians, 
\eqr{eq:fit_gd} and 
\eqr{eq:loop_evar} from the next section.}
%%%%%%%%%%%%%%%%%%%%%%%
\begin{tabular}{|lll|rr|}
\hline
Type & class &pattern & $-\PP_6$ & $|U_{6}|^2$ \\
\hline
$H_2$  & 1 & $(+++---)$ & $0$ &  4 \\
$H_2'$ & 1 & $(+++++-)$ & $0$ &  4 \\
$H_4$  & 2 & $(++--+-)$ & $0$ &  4 \\
$H_6$  & 2 & $(+-+-+-)$ & $-1$ & 36 \\
\hline
\end{tabular}
\label{tab:loops}
\end{table}
%%%%%%%%%%%%%%%%%%%%

Identifying ground states depends on counting the number 
of loops with various spin patterns, since this is what
the effective Hamiltonian depends on.
We first do it for
the shortest loops, starting with hexagons.  
A hexagon that satisfies the $\pi$-flux constraint 
must have one of the four spin patterns shown in 
Table~\ref{tab:loops}); we label the types
``$H_{2m}$'' where $2n$ is the number of AFM bonds in the loop.
Also, independent of the spin pattern, the sites of a hexagon 
are placed in two possible ways within the layer
stacking, which are the ``classes'' explained in the next  paragraphs;
the classes are also labeled in Fig.~\ref{fig:slabs}.

First, there are two classes of hexagon placement 
Class (1) hexagons are centered on thin slabs.  The two spins in the thin layer
are opposite, and each pair within a thick layer is parallel.
Consequently, for each thin slab, the class 1 loops 
are half type $H_2$ and half $H_2'$ (see Table ~\ref{tab:loops}).
Class (2) hexagons span one thick and one thin slab 
The part of the loop within the thick slab, 
always has $+-+-$, so for each thick slab,
the class 2 loops are half type $H_4$ and half type $H_6$, 
of which the last is the type favored by the
effective Hamiltonian.
These are the four hexagon patterns satisfying the $\pi$-flux
condition (\ref{eq:pi_flux}); that confirms that 
these slab stacked states are indeed harmonic ground states, 
a precondition for being hexagon ground states.  
Furthermore, since there are twice as many 
Class 2 hexagons as Class 1, exactly 1/3 of all hexagons are
type H2 (the favored kind). Appendix F of ref.~\onlinecite{uh_thesis} 
shows that a fraction 1/3 is the upper limit, so these are in fact
hexagon ground states, too.
A similar enumeration can  be done  of octagons. Again,  for
each particular type of spin pattern for an octagon, the number 
is the same for all our stacked hexagon ground states,
therefore they are {\it degenerate} up to order 8.

\subsection{Long loops}

Symmetry can be used to show that much longer loops have the
same count in all possible stackings.
Say that a certain loop spans $t$ slabs; the $2^t$ possible
spin states of those slabs are defined by $(s_1, s_2, ..., s_t)$
where each $s_i=\pm 1$ is a reference spin in slab $i$.
Now, a lattice symmetry operation $g$ (which maps each layer to itself)
has the action effect of flipping
the spins in some slabs and not others: i.e. 
$(s_1,s_2,...,s_t)$ is multipled by some pattern of 
$(\gamma_1, \gamma_2, ..., \gamma _t)$ of $\pm 1$ factors, 
depending on $g$.  Provided $t$ is not too large, in fact 
{\it every} possible pattern of $\gamma _i $  is generated
by some one of the lattice symmetries: hence, all stacks
of $t$ slabs are related by symmetry and have the same counts
of all possible loops.  
The smallest stack for which this no longer happens is 
when the first and last slab are stacked directly on top of each other,
which (as worked out above) first happens for $m=4$, meaning
12 layers or for $t=9$ slabs (including the repeated one).  The smallest loop
which requires all of these slabs has length 2(12)+2 = 26.

We conjecture that at order 26, the effective Hamiltonian 
{\it does} break the degeneracy.  That will be a tiny energy:
from \eqr{eq:C2n_100} one could guess $|C_{26}|$ (for $S=100$)
is in the range $10^{-7}$ to $O(10^{-3})$ (depending whether one 
assumes an exponential decrease with $2n$, or a power law).

\section{Ground state problem as coloring}
\label{app:color-grstates}

%TTTTTTTTTTTTTTTTTTTTTTTTTTTTTTTTTTTTTTTTTTTTTTTTTTTTTTTTTTTTTTTTTTTTTT
\begin{table}
\caption{Supertetrahedra types: 
frequencies in hexagon-ground-states,
and the counts of hexagons in each supertetrahedron (using the 
type labels of Table~\ref{tab:loops}.)  The types are given 
color names as explained in text.}
%%%%%%%%%%%%%%%%%%%%%%% 
\begin{tabular}{|llcccc|}
\hline
Type  & Name &  frequency & \multicolumn{3}{c|}{Hexagons}  \\
      &      &            & $H_2$, $H_2$' & $H_4$ & $H_6$\\
      &      &            & orange & white & purple\\
\hline 
a     & purple &  $1/3$  &   0  &  2 &  2 \\
b     &        &    0  &   0  &  3 &  1 \\
c     & orange &  $2/3$  &   2  &  1 &  1 \\
d     &        &    0  &   2  &  2 &  0 \\
\hline
\end{tabular}
\label{tab:supertet}
\end{table} 
%%%%%%%%%%%%%%%%%

Here we consider the ground states of the anharmonic effective
hexagon-order Hamiltonian, $\PP_6$.  We review the arguments
from Appendix F of Ref.~\onlinecite{uh_thesis}.
The key idea is that, in a $\pi$-flux state, there are constraints
on spin arrangements due to the fact that different hexagons 
share edges.  The level at which these contraints are first
important is the {\it super-tetrahedron}, a cluster in the
form of a truncated tetrahedron with four hexagonal faces.
The centers of the super-tetrahedra form the {\it complementary}
diamond lattice 
with the same lattice constant as the diamond lattice
formed by centers of the original tetrahedron lattice.
Each bond of the complementary diamond lattice 
(henceforth ``superbonds'') corresponds 1-to-1 with a 
hexagon in the original pyrochlore lattice.

\SAVE{The bond centers of the complementary diamond lattice
are the complementary pyrochlore lattice, 
used in Refs. \onlinecite{clh_harm} and \onlinecite{uh_harm}.}

We can classify supertetrahedra according  to the types of hexagon
loops appearing on their faces.  Counting arguments there showed 
that there are four classes  (Table \ref{tab:supertet}) and 
the total number of type 6 hexagons is maximized when only 
class (a) and (c) appear.  

\subsection{Octagons in supertetrahedra}
\label{app:octagons}

First we can apply the supertetrahedron enumeration
to show that all the hexagon-ground-states also
are degenerate at the octagon term; we take advantage
of the fact that every octagon is contained entirely
within one supertetrahedron (three contained in each).

We know that any hexagon ground state has fixed fractions
of type (a) and type (b) supertetrahedra, as shown 
in Table~\ref{tab:supertet}.  But each of those supertetrahedra
has a fixed pattern for its octagon loops:
type (a) has one each of $(+-+-+-+-+-)$, $(++-+--+-)$, and
$(++--+-+-)$, while type (b) has one each of
$(++--+-+-)$, $(+++---+-)$, and $(+++++-+-)$.
Hence, any hexagon-ground-state has a fixed
frequency of each octagon loop; from the list just
given and the supertetrahedron frequencies in
Table~\ref{tab:supertet}, the octagon terms have the
values $\PP_8=1/9$, or mean $|U_8|^2 = 64/3$.

\subsection{Node and superbond constraints as coloring rules}

A convenient necessary (though not sufficient) condition to be a
hexagon ground state can be expressed as the following
coloring problem on the complementary diamond lattice.  For this purpose, the
hexagon types (which are the superbonds on this lattice)
are associated with colors, as are the supertetrahedron types
(nodes on the lattice).  Then we have a complete covering by ``purple
trimers'', consisting of two purple bonds (the middle node is purple and
the other two nodes are orange.  Simultaneously, we have a loop covering
by orange loops (connecting orange nodes).
Notice that, if we have such a coloring, we still must verify whether the
can be filled in around each hexagon in a consistent fashion.

In the stacking of Sec.~\ref{sec:groundstates}, 
the supertetrahedra centered in B slabs are of type (a), and those 
centered between A and B slabs are of type (c).
The purple trimer bonds are all oriented vertically (i.e. the three nodes
are always at three different levels); this give $2^2$ degrees of
freedom per $B$ slab, accounting for all the spin entropy.
The orange loops always run horizontally between the A and B slabs 
(perpendicular to the chains of that A slab).

We conjectured, but did not prove, that the
{\it only} hexagon ground states were
the stackings of Sec.~\ref{sec:groundstates}.
The special constraints of the stackings can be expressed, in the color
language, as follows:
%%%%%%%%%%%%%%%%%
\begin{itemize}
\item[](i) If $\alpha,\beta,\gamma,\delta$ are four successive nodes 
connected by orange bonds, then the $(\alpha\beta)$ and $(\gamma\delta)$
are oriented the same.
\item[] (ii) If $\beta$ is orange and $\gamma$ is a purple node,
and $(\alpha\beta$) is the white bond into $\beta$
while $(\gamma\delta)$ is the purple bond out of $\gamma$, then 
$(\alpha\beta)$ is never oriented the same as $(\gamma\delta)$.
\end{itemize}
We do not know if (i) and (ii) follow from the condition of having
only type (a) and (c) super-tetrahedra, and so we do not know
whether any hexagon ground state exists, besides the stacked
family of Sec.~\ref{sec:groundstates},

\bibliography{pyrochlore}

\end{document}